\documentclass[11pt,a4paper]{article}
\usepackage{amsfonts,amssymb,epsfig,amsmath,enumitem,mathtools,MnSymbol}
\usepackage[utf8]{inputenc}
\usepackage{textcomp}
\usepackage{youngtab}
\usepackage{cite}
\usepackage{hyperref,caption,subcaption}
\usepackage{xcolor,tikz}
\usepackage[mathscr]{euscript}

\renewcommand{\thefootnote}{\fnsymbol{footnote}}

\allowdisplaybreaks

\begin{document}

\makeatletter \@addtoreset{equation}{section} \makeatother
\renewcommand{\theequation}{\thesection.\arabic{equation}}
\renewcommand{\thefootnote}{\fnsymbol{footnote}}	

\begin{titlepage}
\begin{center}

\hfill {\tt KIAS-P18034}\\
\hfill {\tt SNUTP18-002}

\vspace{1.5cm}

{\LARGE\bf Wilson surfaces in M5-branes}

\vspace{1.5cm}

{\Large Prarit Agarwal$^1$, Joonho Kim$^2$, Seok Kim$^1$, Antonio Sciarappa$^2$}

\vspace{1cm}

\textit{$^1$Department of Physics and Astronomy \& Center for
Theoretical Physics,\\
Seoul National University, Seoul 08826, Korea.}\\
\vspace{0.3cm}
\textit{$^2$School of Physics, Korea Institute for Advanced Study, Seoul 02455, Korea.}\\

\vspace{1cm}

E-mails: {\tt  agarwalprarit@gmail.com, joonhokim@kias.re.kr, \\
skim@phya.snu.ac.kr, asciara@kias.re.kr}

\end{center}

\vspace{2.5cm}

\begin{abstract}
% \begin{spacing}{1.2}
\normalsize

We study Wilson surface defects in 6d $\mathcal{N}=(2,0)$ superconformal field theories, engineered by semi-infinite M2-branes intersecting M5-branes. Two independent approaches are used to obtain the Wilson surface observables on the $T^2$ inside $\Omega$-deformed $\mathbf{R}^4 \times T^2$. One approach is to compute 5d $\mathcal{N}=1^*$ instanton partition functions in the presence of Wilson lines, where the instanton corrections capture the 6d Kaluza-Klein momentum modes. The other approach is to study the elliptic genera of 2d $\mathcal{N}=(0,4)$ gauge theories, which we propose as describing 6d self-dual strings in the presence of Wilson surface defects. 
%We make a detailed comparison between these two independent computations. They are precisely in agreement for the Wilson surfaces in minuscule representations. However, we find only a partial agreement for non-minuscule representations and comment about it.
We make a detailed comparison between these two independent computations, which precisely agree for Wilson surfaces in minuscule representations; for non-minuscule representations instead we only find partial agreement, due to technical problems which we comment about.

\end{abstract}

\end{titlepage}

\setcounter{tocdepth}{2}
\tableofcontents

\section{Introduction}

Wilson loop operators in gauge theories are among the most important observables one can study. 
They correspond to the phase factor induced from the parallel transport of a charged probe particle  around a loop.
These non-local and gauge~invariant observables can be used to formulate the classical gauge theory itself, encoding some important aspects of the strongly-coupled gauge dynamics. 
For example the expectation value of a Wilson loop, which consists of a pair of semi-infinite straight lines, captures the interaction potential between a quark and an antiquark, becoming an order parameter for quark confinement.

In a 6d superconformal field theory (SCFT) which contains the two-form potential $B_{\mu\nu}$ and the associated supersymmetric strings, the Wilson loop operator of gauge theories can be generalized to the phase factor induced from moving a charged probe string along a spacetime trajectory given by a 2d surface $\Sigma$. This is the Wilson surface operator \cite{Ganor:1996nf} which we will study throughout this paper. It can \emph{formally} be expressed in the following form:
\begin{align}
	\mathcal{W}_{\mathbf{R}}\left[\Sigma\right] = \text{Tr}_{\mathbf{R}}\left(\mathcal{P}\exp{i\int_{\Sigma} d\sigma^{\mu\nu} (B_{\mu\nu} + \cdots) }\right)
\end{align}
where $\cdots$ must be determined from supersymmetric completion for unbroken generators. As we lack a formalism for 6d SCFTs with a non-Abelian self-dual tensor multiplet, the precise meaning of `representation' can only be clarified through string theory realization of the Wilson surface defects. 
For example, an $n$-wound membrane ending on $\Sigma$ realizes the Wilson surface in rank-$n$ symmetric representation, while $n$ membranes ending on $\Sigma$  engineers the rank-$n$ antisymmetric Wilson surface \cite{Chen:2007ir}.

We will particularly focus on the 6d $\mathcal{N} = (2,0)$ SCFT of $A_{N-1}$-type that describes the worldvolume dynamics of $N$ parallel M5-branes. It contains tensionless BPS strings that are induced from open M2-branes suspended between M5-brane pairs \cite{Witten:1995zh,Strominger:1995ac}. We will consider the tensor branch of $(2,0)$ SCFT defined on $\mathbf{R}^4 \times T^2$, i.e. the configuration in which all M5-branes are separated from each other. Such a separation gives the BPS string a non-zero tension proportional to the distance between M5-branes. Alongside these relatively light strings, we will also introduce heavy probe strings induced from semi-infinite M2-branes ending on M5-branes, which engineer the Wilson surface defect. This is a $\frac{1}{2}$-BPS defect preserving only 8 Poincar\'e supercharges.
%, whose `representation' is determined by the ending shape of the semi-infinite M2-branes on M5-branes. 

%The holographic description of the Wilson surface operators was utilized to study the expectation values and the operator product expansion, for the case of the fundamental representation \cite{Maldacena:1998im,Berenstein:1998ij,Corrado:1999pi} and the large-rank antisymmetric and symmetric representations \cite{Lunin:2007ab,Chen:2007ir,Chen:2007zzr,Chen:2008ds}.
The expectation value and operator product expansion of Wilson surface operators in the fundamental representation was studied using holographic techniques in \cite{Maldacena:1998im,Berenstein:1998ij,Corrado:1999pi}. This was generalized to the case of large-rank antisymmetric and symmetric representations in \cite{Lunin:2007ab,Chen:2007ir,Chen:2007zzr,Chen:2008ds}. The field-theoretic calculations of the Wilson surface defects were made through 5d SYM description  of 6d $(2,0)$ SCFTs on $S^5 \times S^1$ \cite{Kim:2012qf,Minahan:2013jwa, Bullimore:2014upa} and $\Omega$-deformed $\mathbf{R}^4 \times T^2$ \cite{Nekrasov:2015wsu,Kim:2016qqs}. An agreement between two computations was found in \cite{Mori:2014tca} for the large-rank symmetric and antisymmetric Wilson surfaces on $T^2 \subset S^5 \times S^1$.

In this paper, we extend the field-theoretic computation to the case of the Wilson surface observables wrapping $T^2$ inside $\Omega$-deformed $\mathbf{R}^4 \times T^2$ \cite{Nekrasov:2015wsu,Kim:2016qqs} using two independent approaches.
One way is to compute 5d $\mathcal{N}=1^*$ instanton partition function in the presence of Wilson loops, where the instanton corrections capture the 6d Kaluza-Klein momentum modes. 
For this purpose, we consider the ADHM quantum mechanics of SYM instantons with Wilson loops \cite{Tong:2014cha} and compute their Witten indices. Specifically for minuscule representations, which were already studied in  \cite{Nekrasov:2015wsu,Kim:2016qqs}, the ADHM indices take a simple form so that one can easily extract the instanton corrected Wilson loop observables out of them.
We also explain the general structure of the ADHM indices and identify the Wilson loops in various other representations beyond the minuscule ones. 
This 5d approach will be discussed in Section~\ref{sec:5dWilson}.
%
% The other way is to study the elliptic genera of 2d $\mathcal{N}=(0,4)$ gauge theories, which we propose as describing 6d BPS strings in the presence of the Wilson surface defect, generalizing \cite{Haghighat:2013gba,Haghighat:2013tka}. 
% The representation of the defect is determined by the shape of the intersection between semi-infinite M2-branes and M5-branes. This approach will be mainly described in Section~\ref{sec:6d}. And also, these two independent computations will be compared in Section~\ref{subsec:5d6dcompare}, where we report the precise agreement for the minuscule ones and only a partial agreement for non-minuscule ones.
%
The other way is to consider the elliptic genera of 2d  $\mathcal{N}=(0,4)$ quiver gauge theories that describe 6d BPS strings in the presence of a Wilson surface defect. We describe these quivers in Section~\ref{sec:6d}, thereby generalizing \cite{Haghighat:2013gba,Haghighat:2013tka}. In Section~\ref{subsec:5d6dcompare}, we compare the results from the above two independent computations and find precise agreement for Wilson surfaces in minuscule representations; %. However, due to technical problems, there seems to be only a partial match between the results for non-minuscule representations. We comment on the origin this mismatch.    
however for non-minuscule representations there seems to be only a partial match. This is due to technical problems on which we comment at the end of the section.

\section{Line operators in 5d $\mathcal{N}=1^*$ gauge theory}
\label{sec:5dWilson}

In this section, we shall study $\frac{1}{2}$-BPS line defects in 5d $\mathcal{N}=1^*$ $U(N)$ gauge theory. 
The string theory realization of 5d maximal SYM is a stack of $N$ coincident D4-branes, which we assume to fill the $x^0, \cdots, x^4$ directions. It preserves $SO(1,4)$ Lorentz symmetry and $SO(5)_R$ R-symmetry that rotates the transverse  $x^5, \cdots, x^9$ directions. We consider a particular configuration where all D4-branes are separated from one another along the $x^9$ axis. This corresponds to the Coulomb branch of 5d maximal SYM in which the vector multiplet scalar $\phi_9$ obtains a non-zero VEV, i.e., $\langle \phi_9 \rangle \neq 0$. The VEV breaks the $U(N)$ gauge symmetry into its Abelian subgroup $U(1)^N$. Also, it preserves only $SO(4)_2$ of the $SO(5)_R$ symmetry group 
that we often decompose into $SO(4)_2 \simeq SU(2)_{2L} \times SU(2)_{2R}$. 

The 5d massive particles preserve the $SO(4)_{1}$ little group of $SO(1,4)$, which can be written as $SO(4)_1 \simeq SU(2)_{1L} \times SU(2)_{1R}$. We denote the doublet indices of $SU(2)_{1L}$,~$SU(2)_{1R}$,~$SU(2)_{2L}$,~$SU(2)_{2R}$ by $\alpha, \dot{\alpha}, a, A$, respectively. The 16 supercharges of 5d maximal SYM are then $Q^{\alpha A}, Q^{\dot{\alpha}A}, Q^{\alpha a}, Q^{\dot{\alpha}a}$. 
There are two types of massive $\frac{1}{2}$-BPS particles in the Coulomb phase: W-bosons and instantons. 
W-bosons are electrically charged objects under $U(1)^N \subset U(N)$, preserving $Q^{\dot{\alpha}A}$ and $Q^{{\alpha}a}$. Their mass is proportional to the Coulomb VEV, i.e., $M =\text{Tr}\, (\langle \phi_9 \rangle  \cdot \Pi)$, where $\Pi$ is the vector of charges with respect to $U(1)^N \subset U(N)$. Self-dual instantons are solitonic objects having the topological $U(1)$ charge $k \equiv \frac{1}{8\pi^2} \int_{\mathbf{R}^4}\text{Tr}(F\wedge F)$, which is a positive integer. They satisfy the self-duality condition $F_{ab} = \frac{1}{2}\epsilon_{abcd}F_{cd}$ on the spatial $\mathbf{R}^4$ and preserve $Q^{\dot{\alpha}A}$ and  $Q^{\dot{\alpha}a}$. The unit instanton mass is ${8\pi^2}/{g_5^2}$ with the gauge coupling  $g_5$. More generally, W-bosons and instantons can form $\frac{1}{4}$-BPS bound states that preserve $Q^{\dot{\alpha}A}$, with mass 
\begin{align}
M = \text{Tr}\, (\langle \phi_9 \rangle  \cdot \Pi) + \frac{8\pi^2 k}{g_5^2}.
\end{align}

The line defects can be introduced by adding $N'$ D4$'$-branes which fill the $x^0, x^5, \cdots, x^8$ directions \cite{Tong:2014cha}. The open string excitations connecting  D4- and D4$'$-branes induce heavy fermionic probe particles in the 5d theory living on the D4 branes. This will give rise to $\frac{1}{2}$-BPS line defects that will be invariant under the SUSY transformation generated by $Q^{\dot{\alpha}A}$ and $Q^{{\alpha}a}$. They couple to the 5d gauge field $\tilde{A}_\mu$ and the vector multiplet scalar $\phi_9$ through
\begin{align}
\mathcal{L}_{1d} = \int dt\ \sum_{i=1}^{N'} \chi_i^\dagger (\partial_t - i \tilde{A}_t + \phi_9 + M_i) \chi^i,
\end{align}
where $M_i$ is the background gauge field for the $U(1)^{N'} \subset U(N')$ flavor symmetry that acts on $\chi_{1}, \,\cdots,\, \chi_{N'}$. The mass parameter $M_i$ of the fermion $\chi_i$ sets the energy scale for its excitation. In this paper we will be interested in the partition function of this 5d SYM coupled to the 1d line defects, as introduced above. For the case of $N'=1$, this was first studied in \cite{Nekrasov:2015wsu,Kim:2016qqs}.
The fugacity variable $e^{-M_i}$ counts the number of $\chi_i$ excitations.
% \begin{align}
% Z_{5d/1d} = \int \mathcal{D}\Phi\mathcal{D}\chi\  e^{-S_{5d}[\Phi] - S_{1d}[\Phi, \chi, m]}.
% \end{align}
Taking the series expansion in $e^{-M_i}$ of the 5d/1d QFT partition function, the individual coefficients correspond to Wilson loop operators in various representations \cite{Gomis:2006sb,Tong:2014cha,Assel:2015oxa}, as will be explained in Section~\ref{subsec:adhm-extra}. In other words, the line defect partition function can be viewed as a generating function of Wilson loops.

\subsection{Review of ADHM quantum mechanics}
\label{subsec:adhmqm}
The low energy dynamics of 5d SYM instantons can be approximated by the 1d $\mathcal{N}=(0,4)$ non-linear sigma model whose target space is the instanton moduli space \cite{Manton:1981mp}. However, the target space has singularities at which the instanton size shrinks to zero. Since the small instanton configurations are beyond the legitimate regime of 5d gauge theory, the UV completion of 5d gauge theory prescribes how to resolve the singularities. In particular, the UV completion based on D-brane engineering of the gauge theory motivates the ADHM construction of instantons, realizing the SYM instantons as D0-branes \cite{Douglas:1995bn}. 

\begin{figure}[t]
        \centering
        \includegraphics[]{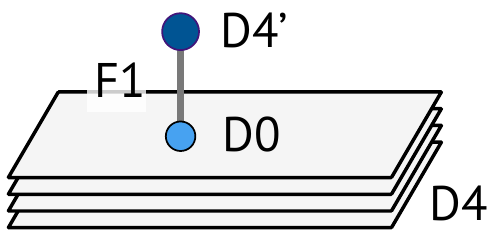}
        \caption{The brane configuration describing 5d instantons in the presence of a 5d Wilson line defect. See figure \ref{fig:WilsonLineQuiver} for the quiver diagram depicting the D0-brane gauged quantum mechanics. }
        \label{fig:WilsonLineBrane}
\end{figure}

The dynamics of 5d SYM instantons in the presence of $\frac{1}{2}$-BPS line defect was  described in \cite{Tong:2014cha}. The insertion of the line defect does not change the static instanton configurations, however, it induces a Lorentz force on slowly moving instantons. 
%Viewing the instantons of 5d maximal SYM as D0-branes sitting on top of D4-branes, 
The ADHM quantum mechanics of 5d instantons in the presence of fermionic line defects is given by the world-volume gauge theory of D0-branes probing the D4-D4$'$ brane configuration (see Figure \ref{fig:WilsonLineBrane}).
The spacetime dimensions occupied by this D0-D4-D4' system are as given in Table~\ref{tbl:brane-d0d4d4'}, where the Taub-NUT space is introduced along the $x^5, \cdots, x^8$ directions to realize the mass deformation of 5d $\mathcal{N}=1^*$ SYM. The UV quantum mechanics corresponding to an instanton configuration with charge $k$ has a $U(k)$ gauge symmetry. It preserves the $SU(2)_{1L} \times SU(2)_{1R} \times SU(2)_{2L} \times SU(2)_{2R}$ global symmetry, where  $\mathcal{N}=(0,4)$ R-symmetry group is $SU(2)_{1R} \times SU(2)_{2R}$. The $U(N)$ and $U(N')$ internal symmetries of D4- and D4$'$-branes also appear as the flavor symmetries of the ADHM quantum mechanics.
\begin{table}[tbp]
	\centering
	\begin{tabular}{c|cccccccccc}
		& 0 & 1 & 2 & 3 & 4 & 5 & 6 & 7 & 8 & 9\\ \hline
		$N$ D4 & $\bullet$ & $\bullet$ & $\bullet$ & $\bullet$ & $\bullet$ &  &  &  &  & \\
		$N'$ D4' & $\bullet$ & & & & & $\bullet$ & $\bullet$ & $\bullet$ & $\bullet$ & \\
		Taub-NUT &  & & & & & $\bullet$ & $\bullet$ & $\bullet$ & $\bullet$ & \\
		$k$ D0 & $\bullet$ & & & & &  &  &  &  & \\
		F1 & $\bullet$ & & & & &  &  &  &  & $\bullet$ \\
	\end{tabular}
	\caption{Brane configuration for 5d $\mathcal{N}=1^*$ $U(N)$ SYM with line defects.}
	\label{tbl:brane-d0d4d4'}
\end{table}

\begin{table}[tbp]
\centering
\begin{subtable}[t]{.49\textwidth}
\vspace{0pt}
\centering
\begin{tabular}{c|c|ccc}
Type & Fields & $U(k)$ & $U(N)$ & $U(N')$ \\\hline
vector & $A_0, \varphi, \zeta_{\dot{\alpha}}^{A}$ & \textbf{adj} & $\mathbf{1}$ & $\mathbf{1}$\\
twisted & $\varphi_{aA},\chi_{a}^{\dot{\alpha}}$ & \textbf{adj} & $\mathbf{1}$ & $\mathbf{1}$\\
hyper & $a_{\alpha\dot{\alpha}}, \lambda_{\alpha}^{A}$ & \textbf{adj} & $\mathbf{1}$ &  $\mathbf{1}$ \\
Fermi & $\lambda^{a}_{\alpha}$ &   \textbf{adj} & $\mathbf{1}$ &  $\mathbf{1}$ \\
hyper & $q_{\dot{\alpha}}, \psi^{A}$ &  $\mathbf{k}$ & $\overline{\mathbf{N}}$  & $\mathbf{1}$ \\
Fermi & $\psi^{a}$ &  $\mathbf{k}$ & $\overline{\mathbf{N}}$  & $\mathbf{1}$ \\
\end{tabular}
\subcaption{from the D0-D0 and D0-D4 string modes}
\label{tbl:1d-field-d0d0d4}
\end{subtable}
\hfill
\begin{subtable}[t]{.49\textwidth}
\vspace{0pt}
\centering
\begin{tabular}{c|c|ccc}
Type & Fields & $U(k)$ & $U(N)$ & $U(N')$ \\\hline
twisted & $\Phi_{A}, \Psi_{\dot{\alpha}}$ & $\mathbf{k}$  & $\mathbf{1}$  & $\overline{\mathbf{N}'}$ \\
Fermi & $\Psi_{\alpha}$ &  $\mathbf{k}$  & $\mathbf{1}$ & $\overline{\mathbf{N}'}$\\
\end{tabular}
\subcaption{from the D0-D4$'$ string modes}
\label{subtbl:1d-field-d0d4p}
\vspace{.68\baselineskip}
\centering
\begin{tabular}{c|c|ccc}
Type & Fields & $U(k)$ & $U(N)$ & $U(N')$ \\\hline
Fermi & $\rho$ & $\mathbf{1}$ &  $\overline{\mathbf{N}}$ & $\mathbf{N}'$\\
\end{tabular}
\subcaption{from the D4-D4$'$ string modes}
\label{subtbl:1d-field-d4d4p}
\end{subtable}
\caption{$\mathcal{N}=(0,4)$ multiplets in the ADHM quantum mechanics.}
\label{tbl:1d-field}
\end{table}

\begin{figure}[t]
        \centering
        \includegraphics[]{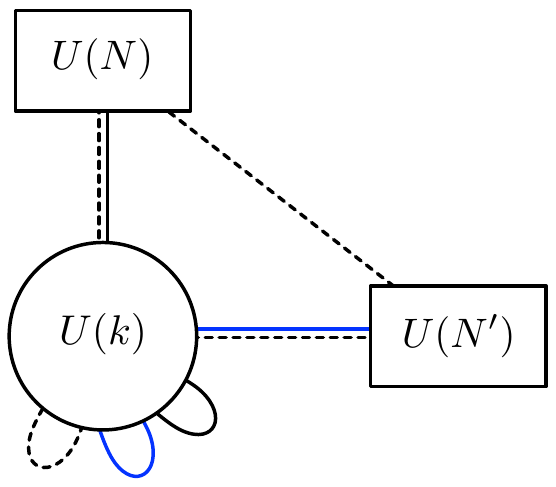}
        \caption{The quiver describing the ADHM quantum mechanics that captures the effect of slow moving 5d instantons in the presence of a Wilson line defect. Here solid black lines denote a hypermultiplet, solid blue lines denote a twisted hyper-multiplet, while the dashed lines denote a Fermi multiplet.  }
        \label{fig:WilsonLineQuiver}
\end{figure}

Various $(0,4)$ multiplets arise from the massless excitations of open strings, contributing to the ADHM quantum mechanics of D0-branes. The D0-D0 strings alone preserve $\mathcal{N}=(8,8)$ supersymmetry and give rise to an $(8,8)$ vector multiplet.
One can decompose the $(8,8)$ vector multiplet into the following $(0,4)$ supermultiplets: a vector multiplet  $(A_0, \varphi, \zeta_{\dot{\alpha}}^{A})$, a twisted hypermultiplet $ (\varphi_{aA},\chi_{a}^{\dot{\alpha}})$, a hypermultiplet $(a_{\alpha\dot{\alpha}}, \lambda_{\alpha}^{A})$, and a Fermi multiplet $(\lambda^{a}_{\alpha})$. All of these transform in the adjoint representation of the $U(k)$ gauge symmetry. The D0-D4 strings provide a $(0,4)$ hypermultiplet $(q_{\dot{\alpha}}, \psi^{A})$ and a Fermi multiplet $(\psi^{a})$, which combine into a $(4,4)$ hypermultiplet transforming in the bifundamental representation of $U(k)\times U(N)$. The above multiplets form the field content of the ADHM quantum mechanics for 5d $\mathcal{N}=1^*$ SYM instantons without line defects. With addition of the line defects associated to $N'$ D4$'$-branes, the massless modes of D0-D4$'$ strings further introduce a $(0,4)$ twisted hypermultiplet $(\Phi_{A}, \Psi_{\dot{\alpha}})$ and a Fermi multiplet $(\Psi_{\alpha})$, transforming in the bifundamental representation of $U(k)\times U(N')$. Finally, the D4-D4' strings also induce non-trivial degrees of freedom to the ADHM quantum mechanics. These correspond to a Fermi multiplet $(\rho)$ transforming in the bifundamental representation of $U(N) \times U(N')$. Although being a singlet of the $U(k)$ gauge symmetry, it plays a crucial role in writing the SUSY invariant interaction.
It is the coupling of this field to the rest, that induces the Lorentz-force on slow moving instantons. The gauged quantum mechanics resulting from the interaction of the above multiplets is shown in Figure \ref{fig:WilsonLineQuiver}.

The Lagrangian of this SUSY quantum mechanics was explicitly constructed in \cite{Tong:2014yna,Tong:2014cha} by employing $\mathcal{N}=(0,2)$ superfield formalism \cite{Witten:1994tz}. For each Fermi multiplet $\eta$, they introduce the  potentials $E_\eta$ and $J_\eta$ as holomorphic functions of $(0,2)$ chiral multiplets in a way that ensures $\mathcal{N}=(0,4)$ enhancement.
We decompose the $(0,4)$ vector and (twisted) hypermultiplets into the $(0,2)$ multiplets as follows: 
\begin{align}
\text{ vector } (A_0, \varphi, \zeta_{\dot{\alpha}}^{A}) &\longrightarrow\ \text{ vector } V\ (A_0, \varphi, \zeta^{1}_{\dot{1}}) \ \oplus\text{ Fermi }\Lambda\ (\zeta^{1}_{\dot{2}}) \nonumber\\
\text{ hyper } (a_{\alpha\dot{\alpha}}, \lambda_{\alpha}^{A})& \longrightarrow \ \text{ chiral } B\ (a_{1\dot{1}}, \lambda_{1}^{2})\  \oplus \text{ chiral }\tilde{B}^{\dagger}\ (a_{1\dot{2}}, \lambda_{1}^{1}) \nonumber\\
\text{ hyper } (q_{\dot{\alpha}}, \psi^{A})& \longrightarrow \ \text{ chiral } q\ (q_{\dot{1}}, \psi^{2})\  \oplus \text{ chiral }\tilde{q}^{\dagger}\ (q_{\dot{2}}, \psi^{1})\\
\text{ twisted hyper } (\Phi_{A}, \Psi_{\dot{\alpha}})& \longrightarrow \ \text{ chiral } \Phi \ (\Phi_{1}, \Psi_{\dot{1}})\  \oplus \text{ chiral }\tilde{\Phi}^{\dagger}\ (\Phi_{2}, \Psi_{\dot{2}})\nonumber\\
\text{ twisted hyper } (\varphi_{aA}, \xi_{a}^{\dot{\alpha}})& \longrightarrow \ \text{ chiral } \varphi\ (\varphi_{11}, \xi_1^{\dot{2}})\  \oplus \text{ chiral }\tilde{\varphi}^{\dagger}\ (\varphi_{12}, \xi_1^{\dot{1}})\nonumber
\end{align}
All $(0,4)$ Fermi multiplets are equivalent to $(0,2)$ Fermi multiplets. 
Following \cite{Tong:2014yna}, we turn on $J_\Lambda$ ($E_\Lambda$) 
as the following holomorphic functions of $(0,2)$ chiral multiplets: 
%from (twisted) hypermultiplets, respectively.
\begin{align}
\label{eq:pot1-adhmqm}
J_\Lambda = q\tilde{q} + [B, \tilde{B}], \qquad 
E_\Lambda = \Phi\tilde{\Phi} + [\varphi, \tilde{\varphi}]
\end{align}
To meet $\mathcal{N}=(0,2)$ SUSY preserving condition, $\sum_\eta \text{Tr} \,(J_\eta \cdot E_\eta) = 0$,
we also introduce the holomorphic potentials for other Fermi multiplets as follows: 
\begin{gather}
E_{\psi^1} = \tilde{\varphi} q,\quad J_{\psi^1} = -\tilde{q}\varphi, \quad 
E_{\psi^2} = \varphi q, \quad J_{\psi^2} = \tilde{q}\tilde{\varphi}, \quad
E_{\Psi_1} = B \Phi, \quad J_{\Psi_1} = \tilde{\Phi} \tilde{B},\quad 
E_{\Psi_2} = -\tilde{B}\Phi, \quad J_{\Psi_2} = \tilde{\Phi}B,\nonumber\\
E_{\lambda_1^1} = [B,\tilde{\varphi}],
\quad J_{\lambda_1^1} = [\tilde{B}, \varphi], \quad 
E_{\lambda_1^2} =  [B, \varphi],\quad
J_{\lambda_1^2} =  -[\tilde{B}, \tilde{\varphi}],\quad E_{\rho} = \tilde{\Phi}q,  \quad J_{\rho} = -\tilde{q}\Phi.
\end{gather}
Combined with the standard D-term $D = q q^{\dagger}-\tilde{q}^{\dagger}\tilde{q} + \Phi\Phi^{\dagger}-\tilde{\Phi}^{\dagger}\tilde{\Phi} + [B,B^{\dagger}] - [\tilde{B}^{\dagger},\tilde{B}] + [\varphi,\varphi^{\dagger}] - [\tilde{\varphi}^{\dagger},\tilde{\varphi}]$, 
the bosonic potential $V = \sum_{\eta} (|E_\eta|^2 + |J_\eta|^2) + \frac{1}{4}D^2 $  can be arranged into 
\begin{align}
\label{eq:pot3-adhmqm}
V &=   \frac{1}{4} \Big((\sigma^m)^{\dot{\alpha}}{}_{\dot{\beta}}  q_{\dot{\alpha}}q^{\dagger\dot{\beta}} + \frac{1}{2}(\sigma^m)^{\dot{\alpha}}{}_{\dot{\beta}} [a_{\alpha \dot{\alpha}}, a^{\alpha\dot{\beta}}] \Big)^2 
+ \frac{1}{4} \Big((\sigma^I)^{A}{}_{B}  \Phi_{A}\Phi^{\dagger B}+\frac{1}{2}(\sigma^I)^{A}{}_{B} [\varphi_{aA}, \varphi^{aB}]\Big)^2 \\
&+ \frac{1}{2}|\Phi^{\dagger A} q_{\dot{\alpha}} |^2 + \frac{1}{2}|\varphi_{ aA}q_{\dot{\alpha}}|^2  + \frac{1}{2} |a_{\alpha\dot{\alpha}}  \Phi_A  |^2 + \frac{1}{4} | a_{\alpha\dot{\alpha}} \varphi_{aA} -  \varphi_{aA}a_{\alpha\dot{\alpha}}|^2.  \nonumber
\end{align}
where $m$ and $I$ are triplet indices for $SU(2)_{1R}$ and $SU(2)_{2R}$, respectively and $\sigma^m ,\, \sigma^I$ are the corresponding Pauli matrices.

Let us examine the classical moduli space of the ADHM quantum mechanics. 
The Higgs branch is parameterized by VEVs of the $SU(2)_{1R}$ doublet fields, $q_{\dot{\alpha}}$ and $a_{\alpha\dot{\alpha}}$, subject to the following ADHM constraint:
\begin{align}
\label{eq:ADHM-constraint}
(\sigma^m)^{\dot{\alpha}}{}_{\dot{\beta}}  q_{\dot{\alpha}}q^{\dagger\dot{\beta}} + \tfrac{1}{2}(\sigma^m)^{\dot{\alpha}}{}_{\dot{\beta}} [a_{\alpha \dot{\alpha}}, a^{\alpha\dot{\beta}}] = 0.
\end{align}
All $\Phi_A$ and $\varphi_{aA}$ fluctuations become massive, except at small instanton singularities where the VEV $\langle q_{\dot{\alpha}} \rangle$ is zero. This shows that the Higgs branch corresponds to the $U(N)$ instanton moduli space. Likewise, the twisted Higgs branch is spanned by VEVs of the $SU(2)_{2R}$ doublet fields, $\Phi_A$ and $\varphi_{aA}$, which are subject to the following constraint having the same functional form as \eqref{eq:ADHM-constraint}:
\begin{align}
(\sigma^I)^{A}{}_{B}  \Phi_{A}\Phi^{\dagger B}+\tfrac{1}{2}(\sigma^I)^{A}{}_{B} [\varphi_{aA}, \varphi^{aB}] = 0.
\end{align}
Moving away from singularities $\langle \Phi_A\rangle = 0$, the development of the $q_{\dot{\alpha}}$ and $a_{\alpha\dot{\alpha}}$ moduli is also suppressed. So the twisted Higgs branch is the $U(N')$ instanton moduli space with exchange of $SO(4)_1 \leftrightarrow SO(4)_2$.
Finally, the mixed branch emerges when $\langle q_{\dot{\alpha}} \rangle = 0$ and $\langle \Phi_A\rangle = 0$, being parameterized by $a_{\alpha\dot{\alpha}}$ and  $\varphi_{aA}$. This corresponds to freely moving D0-branes, unbound to any of background D4- or D4$'$-branes.
We remark that only the Higgs branch is relevant from the 5d SYM perspective. Other branches emerge due to the extra degrees of freedom introduced for the UV completion of 5d SYM instantons. 
Their effects on the Witten index must be properly taken into account, to obtain the correct 5d SYM observables via the ADHM quantum mechanics. 
We discuss this issue in detail in Section~\ref{subsec:adhm-extra}.

\subsection{ADHM index with line operators}
\label{subsec:adhm-index}

We consider the BPS partition function of the 5d/1d coupled system on Omega-deformed $\mathbf{R}^4 \times S^1$, which is given by the following trace formula: 
%\begin{align}
%\label{eq:5d1d-partition-def}
%\textstyle Z_{\rm 5d/1d} = \text{Tr}_{\mathcal{H}_{\rm 5d}}\left[(-1)^F e^{-\beta H} q^k t^{2(J_r - J_R)} u^{2J_l} \mu^{2J_f}    \prod_{a=1}^N w_a^{G_a} \prod_{i=1}^{N'} x_l^{Q_l}\right].
%\end{align}
\begin{align}
\label{eq:5d1d-partition-def}
\textstyle Z_{\rm 5d/1d} = \text{Tr}_{\mathcal{H}_{\rm 5d}}\Big[(-1)^F e^{-\beta H} q^k t^{2(J_r - J_R)} u^{2J_l} \mu^{2J_f}    \displaystyle\prod_{a=1}^N w_a^{G_a} \prod_{i=1}^{N'} x_l^{Q_l}\Big].
\end{align}
It counts the BPS states in 5d $\mathcal{N}=1^*$ SYM, 
annihilated by the supercharges $Q^{1\dot{1}}$ and $Q^{2\dot{2}}$  satisfying $H \sim \{Q^{1\dot{1}}, Q^{2\dot{2}}\}$. 
%The Cartan generators for  $SU(2)_{1L}, SU(2)_{1R}, SU(2)_{2L}, SU(2)_{2R}$ global symmetries are $J_{l}, J_{r}, J_{f}, J_{R}$ respectively. 
Here, $J_{l}, J_{r}, J_{f}, J_{R}$ are the Cartan generators for  $SU(2)_{1L}, SU(2)_{1R}, SU(2)_{2L}, SU(2)_{2R}$ global symmetries respectively.
$G_a$ and $Q_l$ denote the respective Cartan generators for 5d $U(N)$ gauge and $U(N')$ flavor symmetry. $k$~is the topological $U(1)$ charge for 5d SYM instantons which is conjugate to the instanton fugacity $q \equiv \exp\,(-{8\pi^2}/{g_{\rm 5d}^2})$.
For all combinations of the Cartan generators commuting with $Q^{1\dot{1}}$ and $Q^{2\dot{2}}$, the conjugate fugacity variables are introduced as follows:
\begin{align}
t = e^{-\epsilon_+},\ u = e^{-\epsilon_-},\ \mu = e^{-m},\ x_l = e^{-M_l},\ w_a = e^{-\alpha_a}.
\end{align}
$\epsilon_+ \equiv \frac{\epsilon_1 + \epsilon_2}{2}$ and $\epsilon_- \equiv \frac{\epsilon_1 - \epsilon_2}{2}$ introduce the Omega-deformation of the spatial $\mathbf{R}^4$ in the 5d gauge theory. $m$ is the mass of the $\mathcal{N}=1$ adjoint hypermultiplet in the $\mathcal{N}=2$ vector multiplet, realizing the mass deformation of $\mathcal{N}=1^*$ SYM. $\alpha_a$ parametrizes the VEV of the $\mathcal{N}=1$ vector multiplet scalar, i.e., $\langle \phi_9 \rangle = \text{diag}(\alpha_1, \alpha_2, \cdots, \alpha_N)$, that breaks $U(N)$ into $U(1)^N$.
These chemical potentials deform the 5d/1d coupled system by acting as infrared regulators which generate a mass gap for the instanton zero modes \cite{Nekrasov:2002qd,Nekrasov:2003rj}.
$M_l$ is the mass parameter for the 1d defect fermion, breaking the $U(N')$ flavor symmetry into its maximal torus. 
%These chemical potentials deform the 1d/5d coupled system as infrared regulators, which generate the mass gap for the instanton zero modes.\cmt{reference}

The partition function in \eqref{eq:5d1d-partition-def} represents the grand canonical ensemble of all multi-instanton BPS states. One can write the full partition function as the sum over all instanton sectors, multiplied by the perturbative partition function $Z_\text{pert}$ that captures the bound states of W-bosons. 
%Therefore we have to study the sum over the Witten indices of the ADHM quantum mechanics for the full tower of D0-branes.
It implies that we have to sum over the full tower of D0-branes, thereby giving us 
%\begin{align}
%Z_\text{ADHM} \equiv \sum_{k=0}^{\infty} q^k Z_k \ \text{ with } \ 
%\textstyle Z_k \equiv \text{Tr}_{\mathcal{H}_{\rm 1d}^{(k)}}\left[(-1)^F e^{-\beta H} q^k t^{2(J_r - J_R)} u^{2J_l} \mu^{2J_f}    \prod_{a=1}^N w_a^{G_a} \prod_{i=1}^{N'} x_l^{Q_l}\right],
%\end{align}
\begin{align}
\label{eq:ZADHM}
Z_\text{ADHM}^{(N,N')} \equiv \sum_{k=0}^{\infty} q^k Z_k^{(N,N')} \ \text{ with } \ 
\textstyle Z_k^{(N,N')} \equiv \text{Tr}_{\mathcal{H}_{\rm 1d}^{(k)}}\Big[(-1)^F e^{-\beta H}  t^{2(J_r - J_R)} u^{2J_l} \mu^{2J_f}   \displaystyle \prod_{a=1}^N w_a^{G_a} \prod_{i=1}^{N'} x_l^{Q_l}\Big],
\end{align}
where the trace is taken over all the BPS states in the ADHM quantum mechanics of $k$ D0-branes. The corresponding Witten index $Z_k^{(N,N')}$ can be computed by using the results of \cite{Hwang:2014uwa,Hori:2014tda,Cordova:2014oxa}.

%We apply the result of \cite{Hwang:2014uwa,Hori:2014tda,Cordova:2014oxa} to compute the Witten index $Z_k$ of the ADHM quantum mechanics. 
Let us evaluate the gauge theory path integral in the weak coupling regime $g_1 \rightarrow 0$ of the UV quantum mechanics, reducing it down to Gaussian integrals around saddle points. The saddle points are parameterized by the complexified holonomy $\phi = i\beta A_0 + \beta \varphi$, 
made of the gauge field $A_0$ and the scalar $\varphi$ in the vector multiplet.
We denote the eigenvalues of $\phi$ by $\phi_1,\, \cdots,\, \phi_k$, each of which  takes value on a cylinder of radius $\frac{\beta}{2\pi}$. %Performing the Gaussian integrals over massive fluctuations gives us the following 1-loop determinant:
Performing the Gaussian integrals over massive fluctuations we remain with the zero mode integration of the resulting 1-loop determinant:
%\begin{align}
%\label{eq:1loop-adhm}
%Z^\text{1-loop}_{k\,(N,N')} = \frac{1}{k!}\prod_{i=1}^k &\Bigg(\frac{d\phi_i}{2\pi i} \cdot \frac{\prod_{j\neq i}^k \text{sh}(\phi_{ij})\cdot  \prod_{j=1}^k \text{sh}(2\epsilon_+ + \phi_{ij}) \, \text{sh}(m \pm \epsilon_- + \phi_{ij})}{ \prod_{j=1}^k  \text{sh}(\epsilon_+\pm\epsilon_- + \phi_{ij})\text{sh}(-\epsilon_+ \pm m + \phi_{ij})} \\
% &\times \prod_{a=1}^N \frac{\text{sh}(m \pm  (\phi_i - \alpha_a))}{\text{sh}(\epsilon_+ \pm  (\phi_i - \alpha_a))} \prod_{l=1}^{N'} \frac{\text{sh}(+\epsilon_- \pm (\phi_i - M_l))}{\text{sh}(-\epsilon_+ \pm (\phi_i - M_l))} \Bigg)\times \prod_{a=1}^N\prod_{l=1}^{N'} \text{sh}(\alpha_a - M_l),\nonumber
%\end{align}
\begin{align}
\label{eq:1loop-adhm}
Z_{k}^{(N,N')} & = \frac{1}{k!} \oint \left[ \prod_{i=1}^k \frac{d\phi_i}{2\pi i} \right] Z_{k}^{\text{vec}}(\phi, \alpha, \epsilon_{1,2}) Z_{k}^{\text{adj}} (\phi, \alpha, m, \epsilon_{1,2}) \prod_{l = 1}^{N'} Z_{k}^{\text{1d}} (\phi, \alpha, M_l, \epsilon_{1,2}) ,
\end{align}
where (by using $\text{sh}(x)\equiv 2\sinh(x/2)$)
\begin{align}
\label{eq:integrand-adhm}
Z_{k}^{\text{vec}}(\phi, \alpha, \epsilon_{1,2}) & = 
\prod_{\substack{j\neq i \\ i,j = 1}}^k \text{sh}(\phi_{ij}) \prod_{i,j=1}^k \dfrac{\text{sh}(2\epsilon_+ + \phi_{ij})}{\text{sh}(\epsilon_+\pm\epsilon_- + \phi_{ij})}
\prod_{i = 1}^k \prod_{a=1}^N \frac{1}{\text{sh}(\epsilon_+ \pm  (\phi_i - \alpha_a))}, \nonumber \\
Z_{k}^{\text{adj}} (\phi, \alpha, m, \epsilon_{1,2}) & = 
\prod_{i,j=1}^k \dfrac{\text{sh}(m \pm \epsilon_- + \phi_{ij})}{\text{sh}(-\epsilon_+ \pm m + \phi_{ij})}
\prod_{i = 1}^k \prod_{a=1}^N \text{sh}(m \pm  (\phi_i - \alpha_a)),
\\
Z_{k}^{\text{1d}} (\phi, \alpha, M_l, \epsilon_{1,2}) & = 
\prod_{i = 1}^k \frac{\text{sh}(\epsilon_- \pm (\phi_i - M_l))}{\text{sh}(-\epsilon_+ \pm (\phi_i - M_l))}
\prod_{a=1}^N \text{sh}(\alpha_a - M_l). \nonumber
\end{align}
%The next step is to integrate over the zero modes. 
%This integral becomes the contour integral over the space of $\phi$ eigenvalues, where the integral measure \eqref{eq:1loop-adhm} develops simple poles at various locations.
The zero mode integration  translates into a contour integral over the space of $\phi$ eigenvalues. 
% where the integral measure \eqref{eq:1loop-adhm} develops simple poles at various locations. 
The proper choice of the integral contour is done by the Jeffrey-Kirwan residue operation \cite{Hwang:2014uwa,Hori:2014tda,Cordova:2014oxa}. It refers to an auxiliary $k$-dimensional vector $\eta$, although the final residue sum becomes independent of $\eta$. We choose $\eta=(1,\cdots,1)$ to find an explicit expression of the residues.

For the special case of $N'=0$, the Jeffrey-Kirwan residues are labeled by $N$-tuples of Young tableaux $(Y_1, \cdots, Y_N)$ such that $\sum_{a=1}^N \vert Y_a\vert = k$ \cite{Nekrasov:2002qd,Nekrasov:2003rj,Hwang:2014uwa}. We uniquely assign the numbers $1, 2, \cdots, k$ to each of the boxes. Suppose the $i$'th box is at the $m$'th row and $n$'th column of $Y_a$. This represents the following pole location for an integral variable $\phi_i$:
\begin{align}
	\label{eq:YDpole-noD4p}
	\phi_i = \alpha_a - \epsilon_+ - (n-1)\epsilon_1 - (m-1)\epsilon_2.
\end{align}
The sum over all the Jeffrey-Kirwan residues can be written as \cite{Flume:2002az,Bruzzo:2002xf}
\begin{align}
\label{eq:YDformula-noD4p}
Z_{k}^{(N,0)} = \sum_{\sum_{a}|Y_a|=k}\prod_{a,b=1}^N \prod_{s \in Y_a}\frac{\text{sh}(E_{ab}(s)+m-\epsilon_+)\text{sh}(E_{ab}(s)-m-\epsilon_+)}{\text{sh}(E_{ab}(s))\,\text{sh}(E_{ab}(s)-2\epsilon_+)}
\end{align}
where $E_{ab}(s) \equiv \alpha_a - \alpha_b - \epsilon_1 \,h_a(s) + \epsilon_2 \,(v_b(s)+1)$. $v_b(s)$ is the vertical distance from $s$ to the bottom of $Y_b$.
$h_a(s)$ denotes the horizontal distance from $s$ to the right end of $Y_a$.  \eqref{eq:YDformula-noD4p} is the $k$-instanton~correction to the partition function of 5d $\mathcal{N}=1^*$ $U(N)$ SYM without defects.

Similarly, for another special case of $N=0$, the Jeffrey-Kirwan residues are classified by $N'$-tuples of Young tableaux $(Y_1, \cdots, Y_{N'})$ with $\sum_{a=1}^{N'} \vert Y_a\vert = k$ \cite{Nekrasov:2002qd,Nekrasov:2003rj,Hwang:2014uwa}. 
%The pole location and residue sum are almost identical to \eqref{eq:YDpole-noD4p} and \eqref{eq:YDformula-noD4p}, except that the role of $U(N)/U(N')$ and $SO(4)_1$ / SO(4)_2$ symmetries are exchanged with one another, i.e., 
The pole location and residue sum are identical to \eqref{eq:YDpole-noD4p} and \eqref{eq:YDformula-noD4p}, except for the role reversal specified by $U(N) \leftrightarrow U(N')$ and $SO(4)_1 \leftrightarrow SO(4)_2$, i.e.,
\begin{align}
\label{eq:fug-exchange}
(N, \{\alpha_a\} ,\epsilon_+, \epsilon_-, m) \leftrightarrow (N', \{M_l\} ,-\epsilon_+, m, \epsilon_-).
\end{align}
%We note that the multi-particle index $Z_{\rm ADHM}$ of D0-branes is different from the expected partition function $Z_\text{5d/1d}$ for 5d `$U(0)$' SYM coupled to $N'$ defect fermions, i.e., $Z_\text{5d/1d} = 1$ \cite{Hwang:2016gfw}. It is because $Z_k$ captures the bound states of $k$ D0-branes with $n$ D4$'$-branes. In general, the multi-particle index of the ADHM quantum mechanics captures not only the BPS states of the 1d/5d coupled system, but also the extra bound states of D0-branes decoupled from the QFT dynamics. To identify the correct QFT observable, we remove in Section~\ref{subsec:adhm-extra} the extra BPS states emergent after the UV completion of the non-linear sigma model onto the instanton moduli space.
At this point we also note that in the $N=0$ limit, we would have expected the ADHM quantum mechanics to become trivial, i.e. $Z_\text{5d/1d} = 1$. However it turns out that this is not the case for the partition function, $Z_{\rm ADHM}^{(0,N')}$, in the current set-up. This is because when $N=0$, $Z_k^{(0,N')}$ captures the bound states of $k$ D0-branes with the $N'$ D4$'$-branes. In general, the multi-particle index of the ADHM quantum mechanics captures not only the BPS states of the 5d/1d coupled system, but also the extra bound states of D0-branes decoupled from the QFT dynamics on D4-branes. In order to identify the correct QFT observable in Section~\ref{subsec:adhm-extra}, we need to remove  the extra BPS states that arise as a result of the UV completion of the non-linear sigma model onto the instanton moduli space.

We now turn to the general case of $N, N' \neq 0$. All the non-vanishing Jeffrey-Kirwan residues are classified by $(N+N')$-tuples of Young tableaux $\vec{Y} = (Y_1,\cdots,Y_N\, | \, Y_{N+1},\cdots,Y_{N+N'})$ satisfying $\sum_{a=1}^{N+N'} \vert Y_a\vert = k$.  We label all the $k$ boxes by the uniquely assigned integers $1,\cdots,k$. Suppose the $i$'th and $j$'th boxes are located at the $m$'th row and $n$'th column of $Y_{a \leq N}$ and $Y_{l > N}$, respectively. These boxes encode the following information on the pole locations for integral variables $\phi_i$ and $\phi_j$:
\begin{align}
\label{eq:pole-loc-yd}
\phi_i = 
\alpha_a - \epsilon_+ - (n-1)\epsilon_1 - (m-1) \epsilon_2,  \qquad
\phi_j = 
M_l + \epsilon_+ - (n-1)\epsilon_3 - (m-1)\epsilon_4,
\end{align}
with $\epsilon_3 \equiv -\epsilon_+ + m$ and $\epsilon_4 \equiv -\epsilon_+ - m$. 
No poles can be further developed at other locations, such as
$\alpha_a - \epsilon_+ - (n-1)\epsilon_1 - (m-1) \epsilon_2 - \epsilon_{3,4}$ or $M_l + \epsilon_+ - (n-1)\epsilon_3 - (m-1)\epsilon_4 - \epsilon_{1,2}$
due to the suppressing factors
\begin{align*}
 \textstyle \prod_{i=1}^k \left( \prod_{a=1}^N \text{sh}(m \pm  (\phi_i - \alpha_a))\prod_{l=1}^{N'} \text{sh}(+\epsilon_- \pm (\phi_i - M_l)) \prod_{j=1}^k\text{sh}(m \pm \epsilon_- + \phi_{ij})\right)
\end{align*}
in the numerator of \eqref{eq:1loop-adhm}. The JK residue sum can be written as the following expression:
\begin{align}
\label{eq:YDformula-D4p}
&Z_{k}^{(N,N')}= \sum_{\vec{Y}}\Bigg[\prod_{a,b=1}^N \prod_{s \in Y_a}\bigg(\frac{\text{sh}(E_{ab}(s)+m-\epsilon_+)\text{sh}(E_{ab}(s)-m-\epsilon_+)}{\text{sh}(E_{ab}(s))\,\text{sh}(E_{ab}(s)-2\epsilon_+)} \prod_{p=1}^{N'} \frac{\text{sh}(+\epsilon_- \pm (\phi(s) - M_p))}{\text{sh}(-\epsilon_+ \pm (\phi(s) - M_p))} \bigg) \\
&\times  \prod_{p,q=1}^{N'} \prod_{s \in Y_{p+N}}\bigg( \frac{\text{sh}(F_{pq}(s)+\epsilon_-+\epsilon_+)\text{sh}(F_{pq}(s)-\epsilon_-+ \epsilon_+)}{\text{sh}(F_{pq}(s))\,\text{sh}(F_{pq}(s)+2\epsilon_+)} \prod_{a=1}^N \frac{\text{sh}(m \pm  (\phi (s) - \alpha_a))}{\text{sh}(\epsilon_+ \pm  (\phi(s) - \alpha_a))}\bigg) \times \prod_{a=1}^N \prod_{p=1}^{N'} \text{sh}(\alpha_a - M_p) \nonumber\\
&\times \prod_{a=1}^N\prod_{p=1}^{N'} \bigg( \prod_{\substack{s_1 \in Y_a\\s_2 \in Y_{p+N}}} \frac{\text{sh}(\pm(\phi(s_1) - \phi(s_2))) \text{sh}(2\epsilon_+ \pm(\phi(s_1) - \phi(s_2))) \, \text{sh}(m \pm \epsilon_- \pm(\phi(s_1) - \phi(s_2)))}{  \text{sh}(\epsilon_+\pm\epsilon_- \pm(\phi(s_1) - \phi(s_2)))\text{sh}(-\epsilon_+ \pm m \pm(\phi(s_1) - \phi(s_2)))} \bigg)\Bigg],\nonumber
\end{align}
where $F_{pq}(s) \equiv M_p - M_q - \epsilon_3 \,h_p(s) + \epsilon_4 \,(v_q(s)+1)$. $\phi(s)$ denotes the pole location \eqref{eq:pole-loc-yd} for the box~$s$.
We remark that the contribution from the $k=0$ sector is not $1$ but rather $\prod_{a=1}^N \prod_{p=1}^{N'} \text{sh}(\alpha_a - M_p)$, which captures the fermionic excitation of the D4-D4$'$ strings.

Before we conclude the section, let us mention that the partition function $Z_{\text{ADHM}}^{(N,N')}$ of the D0-D4-D4' system can also be interpreted as the $N'$-th qq-character $\mathscr{X}^{U(N)}_{N'}$ of the five-dimensional $\mathcal{N} = 1^*$ $U(N)$ theory introduced in \cite{Nekrasov:2015wsu} (as already noticed in \cite{Kim:2016qqs} for the $N'=1$ case): 
\begin{equation}
Z_{\text{ADHM}}^{(N,N')} \;=\; \mathscr{X}^{U(N)}_{N'}(M_1, \ldots, M_{N'}) ,
\end{equation}
where we made the dependence on the $M_l$ parameters explicit as customary when dealing with qq-characters. 
Of particular importance is the $N'=1$ (aka fundamental) qq-character $\mathscr{X}^{U(N)}_{1}(M_1)$, which can be used to derive Dyson-Schwinger equations for the five-dimensional theory, while the role of $N'>1$ (aka higher) qq-characters is less understood: our analysis in Section \ref{subsec:adhm-extra} will hopefully clarify their meaning a little, by showing that all qq-characters can be interpreted in terms of Wilson loops in arbitrary representations.
An important property of the qq-characters is that they can be written as rational functions of another set of observables for the  5d $\mathcal{N} = 1^*$ $U(N)$ theory known as $\mathscr{Y}$-observables: these can be thought as the instanton-corrected vacuum expectation values of $\mathscr{Y}(M_l) \sim \det(e^{M_l} - e^{\phi})$, with $\phi$ being the combination of the real vector multiplet scalar and the time component of the gauge field. From the ADHM quantum mechanics point of view, the correlator of $n$ such observables (or their inverse) at positions $x_j$ can be computed as \cite{Kim:2016qqs}
\begin{equation}
\prod_{j = 1}^n \mathscr{Y}^{s_j}(x_j) = \sum_{k \geqslant 0} q^k
\dfrac{1}{k!} \oint \left[\prod_{i = 1}^k \dfrac{d \phi_i}{2\pi i} \right] Z^{\text{vec}}_{k} 
Z^{\text{adj}}_{k} \prod_{j = 1}^n \left[ Z^{\text{1d}}_{k}(x_j) \right]^{s_j} ,
\;\;\;\; s_j = \pm 1,
\end{equation}
where in the integrand we have the functions defined in \eqref{eq:integrand-adhm} (although we suppressed the dependence on $\alpha_a,\epsilon_{\pm},m$ for brevity). The contour integral is again evaluated by summing over residues at those poles which are labeled by $N$-tuples of the Young tableaux \eqref{eq:YDpole-noD4p}: this pole prescription then distinguishes the 1-point correlator $\mathscr{Y}(M_1)$ from the fundamental qq-character $\mathscr{X}^{U(N)}_{1}(M_1)$, which is defined by the same contour integral formula but receives contributions from a larger set of poles \eqref{eq:pole-loc-yd}. %As a consequence
In more detail, in terms of $\mathscr{Y}$-observables the fundamental qq-character reads \cite{Nekrasov:2015wsu}
\begin{equation}
\begin{split}
\mathscr{X}_{1}^{U(N)}(M_1) \;= \;& \mathscr{Y}(M_1) \sum_{\lambda} q^{\vert \lambda \vert} 
\prod_{s \in \lambda} S\big( (m-\epsilon_+)(h(s) + 1) + (m + \epsilon_+) v(s) \big) \\
& \prod_{s \in \lambda} \dfrac{\mathscr{Y}(M_1 - m - \epsilon_+ + \sigma(s))\mathscr{Y}(M_1 + m - \epsilon_+ + \sigma(s))}{\mathscr{Y}(M_1 + \sigma(s))\mathscr{Y}(M_1 - 2\epsilon_+ + \sigma(s))} \\
= \; & \mathscr{Y}(M_1) + q S(m-\epsilon_+) \dfrac{\mathscr{Y}(M_1- m - \epsilon_+)\mathscr{Y}(M_1 + m - \epsilon_+)}{\mathscr{Y}(M_1- 2 \epsilon_+)} + \ldots;
\end{split}
\end{equation}
here $\lambda$ is a Young tableau with any number of boxes, $h(s)$ and $v(s)$ are the distances from the position $(i,j)$ of the box $s \in \mathbb \lambda$ to the right and bottom ends of the tableau $\lambda$ respectively, and
\begin{equation}
\sigma(s) = i(m-\epsilon_+) - j(m+\epsilon_+) + 2\epsilon_+\, ,
\qquad S(x) = \dfrac{\text{sh}(x + \epsilon_1)\text{sh}(x + \epsilon_2)}{\text{sh}(x)\text{sh}(x + 2\epsilon_+)}.
\end{equation}
Higher qq-characters can be expressed as rational functions of $\mathscr{Y}$-observables in a similar way:
\begin{equation}
\begin{split}
& \mathscr{X}_{N'}^{U(N)}(M_1, \ldots, M_{N'}) = \\
& \prod_{p = 1}^{N'} \mathscr{Y}(M_p)
\sum_{\vec{\lambda}} q^{\sum_{p=1}^{N'} \vert \lambda_p \vert}
\prod_{p,q = 1}^{N'} \prod_{s \in \lambda_p} S(F_{pq}(s))
\prod_{p = 1}^{N'} \prod_{s \in \lambda_p} \dfrac{\mathscr{Y}(M_p - m - \epsilon_+ + \sigma(s))\mathscr{Y}(M_p + m - \epsilon_+ + \sigma(s))}{\mathscr{Y}(M_p + \sigma(s))\mathscr{Y}(M_p - 2\epsilon_+ + \sigma(s))},
\end{split}
\end{equation}
with $\vec{\lambda} = (\lambda_1, \ldots, \lambda_{N'})$ being an $N'$-tuple of Young tableaux and
\begin{equation}
F_{pq}(s) = M_p - M_q + (m - \epsilon_+)(h_p(s) + 1) + (m+\epsilon_+) v_q(s).
\end{equation}

\subsection{The Wilson loop observable}
\label{subsec:adhm-extra}
Recall that the multi-particle partition function $Z_\text{ADHM}^{(N,N')}$ of D0-branes not only captures the BPS states in the 5d/1d coupled system, but also receives extra contribution depending on the UV completion of the QFT. Since the ADHM prescription is equivalent to the string theory embedding of the 5d/1d QFT, i.e., the D0-D4-D4$'$ brane system, we expect the extra contribution is associated to the bound states of D0- and D4$'$-branes. This is obvious for the simple case of $N=0$, where we get \eqref{eq:YDformula-noD4p} after \eqref{eq:fug-exchange}.
Now we consider $Z_\text{ADHM}$ for a general case and construct the QFT observable out of it.

The multi-particle partition function $Z_\text{ADHM}^{(N,N')}$ starts with $\prod_{a=1}^N\prod_{l=1}^{N'} \text{sh}(\alpha_a - M_l)$ as the $q^0$ correction. Imposing the $SU(N)$ and $SU(N')$ traceless conditions $\sum_{a=1}^N \alpha_a = \sum_{l=1}^{N'} M_l = 0$, it decomposes into the following sum over some irreducible representations $(\mathbf{R}, \mathbf{R}')$ of $SU(N) \times SU(N')$: 
\begin{align}
\label{eq:WilsonLoopZeroq}
Z_{0}^{(N,N')} = \prod_{a=1}^N\prod_{l=1}^{N'} \text{sh}(\alpha_a - M_l) = \sum_{(\mathbf{R},\mathbf{R}')} n_{\mathbf{R},\mathbf{R}'} \cdot \chi_{\mathbf{R}}^{SU(N)}(\alpha_a)\, \chi_{\mathbf{R}'}^{SU(N')}(M_l)
\end{align}
%where $\chi_{\mathbf{R}}^{SU(N)}$ ($\chi_{\mathbf{R}'}^{SU(N')}$) denotes the $SU(N)$ ($SU(N')$) character for the representation $\mathbf{R}$ ($\mathbf{R}'$) and $n_{\mathbf{R},\mathbf{R}'}$ is an integer counting the degeneracy of the representations. 
where $\chi_{\mathbf{R}}^{G}$ denotes the character of the Lie group $G$ in the representation $\mathbf{R}$. Also, $n_{\mathbf{R},\mathbf{R}'}$ is an integer counting the degeneracy of the representations. 
This is also the 5d/1d QFT partition function in \eqref{eq:5d1d-partition-def} at zero instanton order, as the issue of extra bound states does not exist. Henceforth, unless explicitly mentioned, we will always impose the $SU(N)$ and $SU(N')$ traceless conditions $\sum_{a=1}^N \alpha_a = \sum_{l=1}^{N'} M_l = 0$.
%The issue of extra bound states gets taken care of trivially since there are 
%Since the Wilson line expectation values $\mathcal{W}_{\mathbf{R}}$ in a given representation $\mathbf{R}$, without instanton corrections, are given by the $SU(N)$ characters $\chi_{\mathbf{R}}^{SU(N)}$, it agrees with the 1d/5d QFT partition function in \eqref{eq:5d1d-gen}. 

At higher $q$, we expect that, just like the  right most expression in \eqref{eq:WilsonLoopZeroq}, each summand in $Z_\text{ADHM}^{(N,N')}$ will correspond to a product of individual Wilson loops in appropriate representations of $SU(N)$ and $ SU(N')$. The only difference will be that the $SU(N) , \, SU(N')$ characters $\chi_{\mathbf{R}}^{SU(N)}(\alpha_a) , \, \chi_{\mathbf{R}'}^{SU(N')}(M_l)$ will now incur instanton corrections to become $\mathcal{W}_{\mathbf{R}}^{U(N)} , \, \mathcal{W}_{\mathbf{R}'}^{U(N')}$ respectively \footnote{The reason to switch from $SU(N)$ in the  superscript of $\chi_{\mathbf{R}}^{SU(N)}$ to $U(N)$ in the superscript of $\mathcal{W}_{\mathbf{R}}^{U(N)}$ is because even after imposing the $SU(N)$ traceless condition by hand, the $q$ corrected version of $\chi_{\mathbf{R}}^{SU(N)}$ contains a $U(1)$ factor (given explicitly in \eqref{eq:U1Factor}) which needs to be appropriately handled. }: 
\begin{align}
 \label{eq:ADHM-index-factorized}
Z_{\rm ADHM}^{(N,N')} =  \sum_{(\mathbf{R},\mathbf{R}')} n_{\mathbf{R},\mathbf{R}'} \cdot \mathcal{W}_{\mathbf{R}}^{U(N)}\cdot \mathcal{W}_{\mathbf{R}'}^{U(N')}.
\end{align}
We believe that $\mathcal{W}_{\mathbf{R}}^{U(N)}$ captures corrections from the D0-D4 bound states while the $q$ corrections in $\mathcal{W}_{\mathbf{R}'}^{U(N')}$ come from the D0-D4$'$ bound states. This is motivated from the moduli space analysis of section \ref{subsec:adhmqm}, where we show that, in the IR, the Higgs branch and the twisted Higgs branch 
%can be detached by the FI deformations, 
form individual quantum systems decoupled from each other.
% only meet at a point

%For the higher $q$ corrections, the ADHM index $Z_\text{ADHM}^{(N,N')}$ also captures the BPS states decoupled from the 1d/5d QFT system. Based on the decoupling argument of \cite{Hwang:2014uwa} between the (twisted) Higgs branches, we expect that the multi-particle index $Z_\text{ADHM}^{(N,N')}$ is  the sum over $(\mathbf{R}, \mathbf{R}')$ of $SU(N) \times SU(N')$ with summands factorized into the D0-D4 and D0-D4$'$ observables:

The QFT observable $\mathcal{W}_{\mathbf{R}}^{U(N)}$ is the Wilson line of 5d $U(N)$ SYM engineered from $N$ D4-branes, where the effect of the overall $U(1)$ is just to include the 
 $U(1)$ instanton partition function \cite{Iqbal:2008ra,Poghossian:2008ge,Kim:2011mv}:\footnote{The Plethystic exponential is defined as
$\text{PE}[f(q,t,u,\mu,w_a,x_l)] \equiv \exp\left(\sum_{n=1}^\infty \frac{1}{n} f(q^n,t^n,u^n,\mu^n,w_a^n,x_l^n)\right)$.}
\begin{align}
\label{eq:U1Factor}
\mathcal{W}^{U(1)} \equiv \text{PE}\,
\bigg[\frac{t (-u-u^{-1}+\mu+\mu^{-1})}{(1-tu)(1-tu^{-1})} \frac{q}{1-q}\bigg]. 
\end{align}
Similarly, $\mathcal{W}_{\mathbf{R'}}^{U(N')}$ can be interpreted as the Wilson line of 5d $SU(N')$ SYM engineered from $N'$ D4$'$-branes, multiplied by the $U(1)$ factor $\mathcal{W}^{U(1')} \equiv \mathcal{W}^{U(1)}|_{t\rightarrow t^{-1}, u \leftrightarrow \mu}$ for the D4$'$-branes.
If $N=N'$ and $\mathbf{R} = \mathbf{R}'$, the observables $\mathcal{W}_{\mathbf{R}}^{U(N)}$ and $\mathcal{W}_{\mathbf{R}'}^{U(N')} $ have the same functional form up to the fugacity exchange \eqref{eq:fug-exchange}.
Once we have decomposed $Z_{\rm ADHM}^{(N,N')} $ according to \eqref{eq:ADHM-index-factorized}, the 5d/1d QFT partition function \eqref{eq:5d1d-partition-def} can be obtained by suppressing the $q$ correction in $\mathcal{W}_{\mathbf{R}'}^{U(N')}$ as follows:
\begin{align}
\label{eq:part-1d5d}
Z_{\rm 5d/1d} =  \sum_{(\mathbf{R},\mathbf{R}')} n_{\mathbf{R},\mathbf{R}'} \cdot \mathcal{W}_{\mathbf{R}}^{U(N)}\cdot \Big( \mathcal{W}_{\mathbf{R}'}^{U(N')}\big|_{q \rightarrow 0}\Big) = \sum_{(\mathbf{R},\mathbf{R}')} n_{\mathbf{R},\mathbf{R}'} \cdot \mathcal{W}_{\mathbf{R}}^{U(N)}\cdot  \chi_{\mathbf{R}'}^{SU(N')}.
\end{align}

Let us  specialize to the particular case of $N' = 1$ which 
was first studied in \cite{Kim:2016qqs}. 
It is convenient to keep the $U(1)$ flavor fugacity alive, such that $x_1 \neq 1$. The multi-particle index $Z_{\rm ADHM}^{(N,1)}$ is the generating function of the minuscule Wilson lines with the overall dressing factor $\mathcal{W}^{U(1')}$, i.e.,
\begin{align}
\label{eq:ADHMN1}
Z_{\rm ADHM}^{(N,1)} &= (-1)^N x_1^{-N/2} \sum_{i=0}^N (-x_1)^{i} \cdot \mathcal{W}_{\bigwedge^i}^{U(N)} \,  \mathcal{W}^{U(1')},
\end{align}
where the subscript $\bigwedge^n$ denotes the rank-$n$ antisymmetric representation. 
% This observable is also called the fundamental qq-character of 5d $\mathcal{N} = 1^*$ $U(N)$ theory, introduced in \cite{Nekrasov:2015wsu} and further studied in \cite{Bourgine:2017jsi}. One of its importance resides in the observation that Dyson-Schwinger equations for the associated gauge theory can be derived from it. 
We identified all the $SU(N)$ Wilson loops with $N\leq 4$ up to 2 instanton corrections. However, for the sake of brevity, we will only display results up to $q^1$ order. Furthermore, we will normalize the observable by the partition function of $U(N)$ SYM which we denote by $\mathcal{W}_{\mathbf{1}}^{U(N)}$. %which is the same as the partition function of $U(N)$ SYM.
All results are expressed in terms of the irreducible characters $\chi^l_{\bf R},  \chi^f_{\bf R}, \chi^r_{\bf R},\chi^{SU(N)}_{\bf R}$ of the $SU(2)_{1L}, SU(2)_{2L}, SU(2)_r \subset SU(2)_{1R} \times SU(2)_{2R}, SU(N)$ symmetries. 
\begin{align}
\frac{\mathcal{W}_{\mathbf{2}}^{U(2)}}{\mathcal{W}_{\mathbf{1}}^{U(2)}} &= \chi_{\mathbf{2}}^{SU(2)} - q \cdot \frac{t^2\cdot  (\chi^f_{\bf 2}-\chi ^l_{\bf 2})(\chi^f_{\bf 2}-\chi ^r_{\bf 2})}{(1-t^2w_1^{\pm 2})}\cdot \chi_{\mathbf{2}}^{SU(2)} + \mathcal{O}(q^2) \\
\frac{\mathcal{W}_{\mathbf{3}}^{U(3)}}{\mathcal{W}_{\mathbf{1}}^{U(3)}} &= \chi_{\mathbf{3}}^{SU(3)} - q \cdot \frac{t^6 \cdot  (\chi^f_{\bf 2}-\chi ^l_{\bf 2})(\chi^f_{\bf 2}-\chi ^r_{\bf 2})}{\prod_{i\neq j}(1-t^2w_iw_j^{-1})} \Big(
\chi^{SU(3)}_{\bf 24} + (2 - \chi^r_{\bf 3})\chi^{SU(3)}_{\bf 15}    -(\chi^f_{\bf 2} \chi^r_{\bf 2}  + 2 \chi^r_{\bf 3} - 1) \chi^{SU(3)}_{\bf \overline{6}}\nonumber\\
&+(\chi^f_{\bf 2} \chi^r_{\bf 4} + \chi^r_{\bf 5} + 2)\chi^{SU(3)}_{\bf 3}\Big) + \mathcal{O}(q^2) \\
\frac{\mathcal{W}_{\mathbf{4}}^{U(4)}}{\mathcal{W}_{\mathbf{1}}^{U(4)}} &= \chi_{\mathbf{4}}^{SU(4)} + q \cdot \frac{t^{12} \cdot  (\chi^f_{\bf 2}-\chi ^l_{\bf 2})(\chi^f_{\bf 2}-\chi ^r_{\bf 2})}{\prod_{i\neq j}(1-t^2w_iw_j^{-1})} \Big(\chi^{SU(4)}_{\bf 756} + (2-\chi^r_{\bf 3})\chi^{SU(4)}_{\bf 540}+(-2 \chi^r_{\bf 3}+\chi^r_{\bf 5}+4)\chi^{SU(4)}_{\bf 360} -\chi^r_{\bf 3} \chi^{SU(4)}_{\bf 280'}\nonumber
\\
&+(-\chi^f_{\bf 2} \chi^r_{\bf 2}-2 \chi^r_{\bf 3}+1)\chi^{SU(4)}_{\bf 224} +(-\chi^f_{\bf 2} \chi^r_{\bf 2}-3 \chi^r_{\bf 3}+\chi^r_{\bf 5}+2)\chi^{SU(4)}_{\bf 160} +(\chi^f_{\bf 2} \chi^r_{\bf 4}-2 \chi^r_{\bf 3}+2 \chi^r_{\bf 5}+3)\chi^{SU(4)}_{\bf 140'} \nonumber\\&
+(-\chi^f_{\bf 2} \chi^r_{\bf 2}+2 \chi^f_{\bf 2} \chi^r_{\bf 4}-\chi^f_{\bf 2} \chi^r_{\bf 6}+\chi^f_{\bf 3}-5 \chi^r_{\bf 3}+3 \chi^r_{\bf 5}-\chi^r_{\bf 7}+7)\chi^{SU(4)}_{\bf 140} + (\chi^f_{\bf 2} \chi^r_{\bf 4}+\chi^r_{\bf 5}+2)\chi^{SU(4)}_{\bf 120} \nonumber\\&
+(\chi^f_{\bf 2} \chi^r_{\bf 4}-\chi^r_{\bf 3}+\chi^r_{\bf 5}-\chi^r_{\bf 7}+3)\chi^{SU(4)}_{\bf 84'} + (-2 \chi^f_{\bf 2} \chi^r_{\bf 2}-\chi^f_{\bf 3} \chi^r_{\bf 3}-2 \chi^f_{\bf 2} \chi^r_{\bf 6}-5 \chi^r_{\bf 3}+\chi^r_{\bf 5}-2 \chi^r_{\bf 7}+5)\chi^{SU(4)}_{\bf 60}\nonumber\\&
+(-\chi^f_{\bf 2} \chi^r_{\bf 2}+2 \chi^f_{\bf 2} \chi^r_{\bf 4}-\chi^f_{\bf 2} \chi^r_{\bf 6}+\chi^f_{\bf 2} \chi^r_{\bf 8}+\chi^f_{\bf 3}-4 \chi^r_{\bf 3}+3 \chi^r_{\bf 5}+\chi^r_{\bf 9}+5)\chi^{SU(4)}_{\bf 36}
+(-2 \chi^f_{\bf 2} \chi^r_{\bf 2}-\chi^f_{\bf 3} \chi^r_{\bf 3}-\chi^f_{\bf 2} \chi^r_{\bf 4}\nonumber\\&
-\chi^f_{\bf 3} \chi^r_{\bf 5}-2 \chi^f_{\bf 2} \chi^r_{\bf 6}-3 \chi^r_{\bf 3}-\chi^r_{\bf 5}-2 \chi^r_{\bf 7}+2)\chi^{SU(4)}_{\bf 20''}+(\chi^f_{\bf 3} \chi^r_{\bf 5}+\chi^f_{\bf 3} \chi^r_{\bf 7}+4 \chi^f_{\bf 2} \chi^r_{\bf 4}+2 \chi^f_{\bf 2} \chi^r_{\bf 8}+\chi^f_{\bf 3}-2 \chi^r_{\bf 3}+4 \chi^r_{\bf 5}\nonumber\\&
+\chi^r_{\bf 9}+6)\chi^{SU(4)}_{\bf 20}
+(-2 \chi^f_{\bf 2} \chi^r_{\bf 2}-\chi^f_{\bf 3} \chi^r_{\bf 3}-\chi^f_{\bf 2} \chi^r_{\bf 4}-\chi^f_{\bf 3} \chi^r_{\bf 5}-3 \chi^f_{\bf 2} \chi^r_{\bf 6}-\chi^f_{\bf 3} \chi^r_{\bf 7}-\chi^f_{\bf 2} \chi^r_{\bf 8}-\chi^f_{\bf 3} \chi^r_{\bf 9}-\chi^f_{\bf 2} \chi^r_{\bf 10}\nonumber\\&
-4 \chi^r_{\bf 3}-\chi^r_{\bf 5}-2 \chi^r_{\bf 7}-\chi^r_{\bf 9}-\chi ^r{}_{11}+1)\chi^{SU(4)}_{\bf 4}
 \Big) + \mathcal{O}(q^2)\\
 \frac{\mathcal{W}_{\mathbf{6}}^{U(4)}}{\mathcal{W}_{\mathbf{1}}^{U(4)}} &= \chi_{\mathbf{6}}^{SU(4)} + q \cdot \frac{t^{12} \cdot  (\chi^f_{\bf 2}-\chi ^l_{\bf 2})(\chi^f_{\bf 2}-\chi ^r_{\bf 2})}{\prod_{i\neq j}(1-t^2w_iw_j^{-1})} \Big(
 \chi^{SU(4)}_{ \bf 960} + (2-\chi^r_{\bf 3 }) (\chi^{SU(4)}_{ \bf 630}+ \chi^{SU(4)}_{  \overline{\bf 630}}) +(\chi^r_{\bf 5 }-3 \chi^r_{\bf 3 }+4)\chi^{SU(4)}_{ \bf 384} \nonumber \\&
+(\chi^f_{\bf 3 }-5 \chi^r_{\bf 3 }+2 \chi^f_{\bf 2 } \chi^r_{\bf 4 }+4 \chi^r_{\bf 5 }+10 )\chi^{SU(4)}_{ \bf 300} - (\chi^f_{\bf 2 } \chi^r_{\bf 2 }+2 \chi^r_{\bf 3 }-1)(\chi^{SU(4)}_{ \bf 270} + \chi^{SU(4)}_{ \bf \overline{270}}) + (\chi^f_{\bf 2 } \chi^r_{\bf 4 }+\chi^r_{\bf 5 }+2) \nonumber\\& \times
(\chi^{SU(4)}_{ \bf 140''} + \chi^{SU(4)}_{  \overline{\bf 140}''} )-(3 \chi^f_{\bf 2 } \chi^r_{\bf 2 }+\chi^f_{\bf 3 } \chi^r_{\bf 3 }+7 \chi^r_{\bf 3 }-2 \chi^r_{\bf 5 }+\chi^f_{\bf 2 } \chi^r_{\bf 6 }+\chi^r_{\bf 7 }-6)(\chi^{SU(4)}_{ \bf 126} + \chi^{SU(4)}_{ \bf \overline{126}}) + (\chi^f_{\bf 3 } \nonumber\\&
-\chi^f_{\bf 2 } \chi^r_{\bf 2 }-5 \chi^r_{\bf 3 }+2 \chi^f_{\bf 2 } \chi^r_{\bf 4 }+2 \chi^r_{\bf 5 }-2 \chi^f_{\bf 2 } \chi^r_{\bf 6 }-2 \chi^r_{\bf 7 }+7) ( \chi^{SU(4)}_{ \bf 70} + \chi^{SU(4)}_{ \bf \overline{70}}) + (-2 \chi^r_{\bf 3 } \chi^f_{\bf 3 }+2 \chi^f_{\bf 3 }-4 \chi^f_{\bf 2 } \chi^r_{\bf 2 } \nonumber\\&
-12 \chi^r_{\bf 3 }+4 \chi^f_{\bf 2 } \chi^r_{\bf 4 }+6 \chi^r_{\bf 5 }-4 \chi^f_{\bf 2 } \chi^r_{\bf 6 }-3 \chi^r_{\bf 7 }+2 \chi^f_{\bf 2 } \chi^r_{\bf 8 }+\chi^r_{\bf 9 }+15)\chi^{SU(4)}_{ \bf 64} + (\chi^r_{\bf 3 } \chi^f_{\bf 3 }+\chi^r_{\bf 5 } \chi^f_{\bf 3 }+\chi^f_{\bf 3 }+2 \chi^f_{\bf 2 } \chi^r_{\bf 2 } \nonumber\\&
-\chi^r_{\bf 3 }+4 \chi^f_{\bf 2 } \chi^r_{\bf 4 }+4 \chi^r_{\bf 5 }-\chi^r_{\bf 7 }+7)\chi^{SU(4)}_{ \bf 50} + (\chi^r_{\bf 5 } \chi^f_{\bf 3 }+\chi^r_{\bf 7 } \chi^f_{\bf 3 }+\chi^f_{\bf 3 }-\chi^f_{\bf 2 } \chi^r_{\bf 2 }-3 \chi^r_{\bf 3 }+4 \chi^f_{\bf 2 } \chi^r_{\bf 4 }+5 \chi^r_{\bf 5 }+\chi^f_{\bf 2 } \chi^r_{\bf 6 }\nonumber\\&
+2 \chi^r_{\bf 7 }+3 \chi^f_{\bf 2 } \chi^r_{\bf 8 }+2 \chi^r_{\bf 9 }+5)(\chi^{SU(4)}_{ \bf 10}+\chi^{SU(4)}_{ \bf \overline{10}}) - (5 \chi^r_{\bf 3 }+\chi^r_{\bf 5 }+5 \chi^r_{\bf 7 }+ 2 \chi^r_{\bf 9 }+\chi^f_{\bf 3 } (\chi^r_{\bf 3 }+\chi^r_{\bf 5 }+\chi^r_{\bf 7 }+\chi^r_{\bf 9 }-1)\nonumber\\&
+2 \chi^f_{\bf 2 } (\chi^r_{\bf 2 }+3 \chi^r_{\bf 6 }+\chi^r_{\bf 8 }+\chi^r_{\bf 10}) + \chi^r_{\bf 11}-5)\chi^{SU(4)}_{ \bf  6 }
 \Big) + \mathcal{O}(q^2) .
\end{align}
The Wilson loops for all the other minuscule representations can be found from the above ones, by using the rule $\mathcal{W}_{\bigwedge^p}^{U(N)} = \mathcal{W}_{\bigwedge^{N-p}}^{U(N)}(w_i \rightarrow w_i^{-1})$, i.e. by conjugating the $U(1)^N$ electric charges.
Upon suppressing the $q$ correction for the bulk factor $\mathcal{W}^{U(1')}$ in \eqref{eq:ADHMN1}, one can finally obtain the 5d/1d QFT partition functions as follows.
\begin{align}
&Z_{\rm 5d/1d}^{(2,1)} = (x_1 + x_1^{-1}) \mathcal{W}_{\mathbf{1}}^{U(2)} - x_1^{0}  \mathcal{W}_{\mathbf{2}}^{U(2)}, \qquad 
Z_{\rm 5d/1d}^{(3,1)} = (x_1^{3/2} - x_1^{-3/2}) \mathcal{W}_{\mathbf{1}}^{U(3)} + x_1^{-1/2} \mathcal{W}_{\overline{\mathbf{3}}}^{U(3)} - x_1^{1/2} \mathcal{W}_{\mathbf{3}}^{U(3)}, \nonumber\\
&Z_{\rm 5d/1d}^{(4,1)} =  (x_1^{-2} + x_1^{2}) \mathcal{W}_{\mathbf{1}}^{U(4)}- x_1^{-1}\mathcal{W}_{\overline{\mathbf{4}}}^{U(4)}+ x_1^{0} \mathcal{W}_{\mathbf{6}}^{U(4)}   - x_1^{1} \mathcal{W}_{\mathbf{4}}^{U(4)} .
\end{align}

Multiple D4$'$-branes are needed for Wilson loops in more general representations. 
For example, the $U(2)$ symmetric Wilson loop $\mathcal{W}_{\mathbf{3}}^{U(2)}$ appears in the case of $N=N'=2$, for which we expect 
\begin{align}
\label{eq:ADHM22}
	Z_{\rm ADHM}^{(2,2)} &=   -2 \mathcal{W}_{\mathbf{2}}^{U(2)}\, \mathcal{W}_{\mathbf{2}}^{U(2')} + \mathcal{W}_{\mathbf{1}}^{U(2)}\, (\mathcal{W}_{\mathbf{3}}^{U(2')}+\mathcal{W}_{\mathbf{1}}^{U(2')}) + (\mathcal{W}_{\mathbf{3}}^{U(2)}+\mathcal{W}_{\mathbf{1}}^{U(2)})\, \mathcal{W}_{\mathbf{1}}^{U(2')}.
\end{align}
Substituting the previously evaluated $\mathcal{W}_{\mathbf{2}}^{U(2)}$ and its bulk analogue, the combination $\mathcal{W}_{\mathbf{3}}^{U(2)} \mathcal{W}_{\mathbf{1}}^{U(2')} +\mathcal{W}_{\mathbf{1}}^{U(2)}\, \mathcal{W}_{\mathbf{3}}^{U(2')}$ can be found. The two summands in this combination can be disentangled from each other by using the following observations which will also be useful for similar issues that will arise with generic $N$ and $N'$:\footnote{These observations are based on the $SU(2)$, $SU(3)$, $SU(4)$ minuscule Wilson loops with 2 instanton corrections.}
%As in this case, one typically finds the combinations of  $U(N)$ and $U(N')$ Wilson lines in different representations from the ADHM indices \eqref{eq:ZADHM} for the general $N$ and $N'$. Isolation of the individual Wilson lines can be achieved from the following observations:
\begin{itemize}
	\item The $k$-instanton correction of the Wilson lines takes the form of \cite{DelZotto:2016pvm,DelZotto:2017mee,Kim:2018gak}:
	\begin{align}
	\mathcal{W}_{\mathbf{R}}^{U(N)}|_{q^k} = \frac{\sum_{\mathbf{r} } c_{\bf r}(\epsilon_{1,2}, m)\cdot \chi^{SU(N)}_{\bf r}(\alpha_a) }{\prod_{n=1}^k \text{sh}(n\epsilon_1)\,\text{sh}(n\epsilon_2)\cdot \prod^{nm \leq k}_{n,m>0} \prod_{a \neq b}\text{sh}(n\epsilon_1 + m\epsilon_2 + \alpha_a - \alpha_b)}
	\end{align}
	where $\mathbf{r}$ runs over all $SU(N)$ irreducible representations that appear in $\mathbf{R} \otimes \mathbf{adj}^{\ell(N,k)}$. The positive integer ${\ell(N,k)}$ can be determined by inspecting the $k$-instanton partition function $\mathcal{W}_{\mathbf{1}}^{U(N)}|_{q^k}$.
	\item  $\mathcal{W}_{\mathbf{R}}^{U(N)}$ and $\mathcal{W}_{\mathbf{R}'}^{U(N')} $ with $N=N'$ and $\mathbf{R} = \mathbf{R}'$ must have the same functional form up to the fugacity exchange \eqref{eq:fug-exchange}. 

	\item The series expansion of the normalized Wilson line $\mathcal{W}_{\mathbf{R}}^{U(N)} / \mathcal{W}_{\mathbf{1}}^{U(N)} |_{q^k}$ in the $SU(N)$ simple roots $\{\mathfrak{n}_i \equiv w_i/w_{i+1}\,|\, 1 \leq i \leq N-1\}$  starts with the  `leading' states having the following form:
    \begin{align}
    \label{eq:state-leading}
    	\prod_{i=1}^{N-1}\mathfrak{n}_i^{l_i} \quad \text{ with } \quad (l_1,l_2,\cdots,l_{N-1}) = (\lambda_1,\lambda_2,\cdots,\lambda_{N-1}) + (\beta_1,\beta_2,\cdots,\beta_{N-1}) \in \mathbf{R}
    \end{align}
    where $(\lambda_1,\lambda_2,\cdots,\lambda_{N-1})$ is the lowest weight of the representation $\mathbf{R}$ in $\alpha$-basis, $(\beta_1,\beta_2,\cdots,\beta_{N-1})$ can be any $SU(N)$ positive (\emph{not} just simple) root. The coefficients of these leading states carry $SU(2)_r \subset SU(2)_{1R} \times SU(2)_{2R}$ spin, which is bounded as $2j_r \leq k$. More generally, the coefficients of the `subleading' states, having the form of $\prod_{i=1}^{N-1}\mathfrak{n}_i^{l_i}  \cdot \mathfrak{n}_j^\ell$ and not overlapping with the set of `leading' states \eqref{eq:state-leading}, are 
    subject to the following $SU(2)_r$ spin bound condition:
	\begin{align}
		2j_r \leq k + (k+1)\ell.
	\end{align}
\end{itemize}
We utilized the above observations and computed the Wilson loop expectation value $\mathcal{W}_{\mathbf{3}}^{U(2)}$ up to 2 instanton corrections. For brevity, we explicitly show $\mathcal{W}_{\mathbf{3}}^{U(2)}/\mathcal{W}_{\mathbf{1}}^{U(2)}$ only up to the 1-instanton order:
\begin{align}
\frac{\mathcal{W}_{\mathbf{3}}^{U(2)}}{\mathcal{W}_{\mathbf{1}}^{U(2)}} &= \chi_{\mathbf{3}}^{SU(2)} + q \cdot \frac{t^2\cdot  (\chi^f_{\bf 2}-\chi ^l_{\bf 2})}{(1-t^2w_1^{\pm 2})}\cdot \Big(\chi^{SU(2)}_{\bf 3} (-\chi^f_{\bf 2}+\chi^l_{\bf 2}+\chi^r_{\bf 2})
-\chi^f_{\bf 2} \chi^l_{\bf 2} \chi^r_{\bf 2}-\chi^f_{\bf 2}+\chi^l_{\bf 2}+\chi^r_{\bf 2}\Big) + \mathcal{O}(q^2).
\end{align}
The same procedure can be repeated for $(N,N')=(2,3)$ and $(N,N')=(3,2)$, for which we have
\begin{align}
\label{eq:ZADHM32}
Z_{\rm ADHM}^{(3,2)} &= (\mathcal{W}_{\mathbf{6}}^{U(3)}+\mathcal{W}_{\overline{\mathbf{6}}}^{U(3)} +2\mathcal{W}_{\mathbf{1}}^{U(3)})\mathcal{W}_{\mathbf{1}}^{U(2')}  -(\mathcal{W}_{\mathbf{8}}^{U(3)}+\mathcal{W}_{\mathbf{3}}^{U(3)}+\mathcal{W}_{\overline{\mathbf{3}}}^{U(3)})\, \mathcal{W}_{\mathbf{2}}^{U(2')}\nonumber\\&
+(\mathcal{W}_{\mathbf{3}}^{U(3)}+\mathcal{W}_{\overline{\mathbf{3}}}^{U(3)})\, \mathcal{W}_{\mathbf{3}}^{U(2')}  -\mathcal{W}_{\mathbf{1}}^{U(3)}\, \mathcal{W}_{\mathbf{4}}^{U(2')}, \\
Z_{\rm ADHM}^{(2,3)} &= \mathcal{W}_{\mathbf{1}}^{U(2)} (\mathcal{W}_{\mathbf{6}}^{U(3')}+\mathcal{W}_{\overline{\mathbf{6}}}^{U(3')} +2\mathcal{W}_{\mathbf{1}}^{U(3')}) -\mathcal{W}_{\mathbf{2}}^{U(2)} (\mathcal{W}_{\mathbf{8}}^{U(3')}+\mathcal{W}_{\mathbf{3}}^{U(3')}+\mathcal{W}_{\overline{\mathbf{3}}}^{U(3')})\nonumber\\&
+\mathcal{W}_{\mathbf{3}}^{U(2)}(\mathcal{W}_{\mathbf{3}}^{U(3')}+\mathcal{W}_{\overline{\mathbf{3}}}^{U(3')})\,   -\mathcal{W}_{\mathbf{4}}^{U(2)}\mathcal{W}_{\mathbf{1}}^{U(3')}.
\end{align}
Similarly we found the Wilson lines $\mathcal{W}_{\mathbf{4}}^{U(2)}$, $\mathcal{W}_{\mathbf{6}}^{U(3)}$, $\mathcal{W}_{\overline{\mathbf{6}}}^{U(3)} = \left(\mathcal{W}_{\mathbf{6}}^{U(3)}\right)^*$, $\mathcal{W}_{\bf 8}^{U(3)}$ up to 1 instanton order.
\begin{align}
	\frac{\mathcal{W}_{\mathbf{4}}^{U(2)}}{\mathcal{W}_{\mathbf{1}}^{U(2)}} &= \chi_{\mathbf{4}}^{SU(3)} - q \cdot \frac{t^2 \cdot  (\chi^f_{\bf 2}-\chi ^l_{\bf 2})}{ (1-t^2w_1^{\pm 2})} \Big(
\chi^{SU(2)}_{\bf 4} (\chi^f_{\bf 2}+\chi^l_{\bf 2}+\chi^r_{\bf 2})
\nonumber\\&
+ \chi^{SU(2)}_{\bf 2} ( -2 \chi^r_{\bf 4}+\chi^f_{\bf 2} \chi^l_{\bf 2}\chi^r_{\bf 2}-\chi^l_{\bf 3}\chi^r_{\bf 2}-2 \chi^l_{\bf 2} \chi^r_{\bf 3}+\chi^f_{\bf 2} \chi^l_{\bf 3}-\chi^r_{\bf 2}+\chi^f_{\bf 2})
\Big) + \mathcal{O}(q^2) \\
\frac{\mathcal{W}_{\mathbf{6}}^{U(3)}}{\mathcal{W}_{\mathbf{1}}^{U(3)}} &= \chi_{\mathbf{6}}^{SU(3)} + q \cdot \frac{t^6 \cdot  (\chi^f_{\bf 2}-\chi ^l_{\bf 2})}{\prod_{i\neq j}^3 (1-t^2w_iw_j^{-1})} \Big(
(\chi^l_{\bf 2}+\chi^r_{\bf 2}-\chi^f_{\bf 2})\chi^{SU(3)}_{\bf \overline{42}} + (\chi^f_{\bf 2} \chi^r_{\bf 3}-2 \chi^f_{\bf 2}-\chi^l_{\bf 2} \chi^r_{\bf 3}+\chi^l_{\bf 2}+\chi^r_{\bf 2}-\chi^r_{\bf 4})\chi^{SU(3)}_{\bf \overline{24}}\nonumber\\&  + (\chi^r_{\bf 2}-\chi^l_{\bf 2} \chi^f_{\bf 2}  \chi^r_{\bf 2}-\chi^f_{\bf 2}+\chi^l_{\bf 2})\chi^{SU(3)}_{\bf \overline{15}'} + (\chi^f_{\bf 2} \chi^l_{\bf 2} \chi^r_{\bf 4}-\chi^l_{\bf 2}\chi^f_{\bf 2} \chi^r_{\bf 2}+\chi^f_{\bf 3} \chi^l_{\bf 2}+2 \chi^f_{\bf 2} \chi^r_{\bf 3}+\chi^f_{\bf 3} \chi^r_{\bf 2}-5 \chi^f_{\bf 2}-2 \chi^l_{\bf 2} \chi^r_{\bf 3}\nonumber\\&
+3 \chi^l_{\bf 2}+2 \chi^r_{\bf 2}-3 \chi^r_{\bf 4})\chi^{SU(3)}_{\bf \overline{15}} + (-\chi^l_{\bf 2} \chi^f_{\bf 2} \chi^r_{\bf 2}+\chi^f_{\bf 2} \chi^l_{\bf 2} \chi^r_{\bf 4}-\chi^f_{\bf 3} \chi^l_{\bf 2} \chi^r_{\bf 3}+2 \chi^f_{\bf 2} \chi^r_{\bf 3}-\chi^f_{\bf 3} \chi^r_{\bf 4}-4 \chi^f_{\bf 2}+\chi^l_{\bf 2} \chi^r_{\bf 5}+3 \chi^l_{\bf 2}\nonumber\\&+3 \chi^r_{\bf 2}-\chi^r_{\bf 4}+\chi^r_{\bf 6})\chi^{SU(3)}_{\bf 6} + 
(-\chi^l_{\bf 2} \chi^r_{\bf 2}+\chi^f_{\bf 2} \chi^l_{\bf 2} \chi^r_{\bf 4}-\chi^f_{\bf 2} \chi^l_{\bf 2} \chi^r_{\bf 6}+\chi^f_{\bf 3} \chi^l_{\bf 2}+2 \chi^f_{\bf 2} \chi^r_{\bf 3}+\chi^f_{\bf 3} \chi^r_{\bf 2}-\chi^f_{\bf 3} \chi^r_{\bf 4}-4 \chi^f_{\bf 2}\nonumber\\&
-2 \chi^l_{\bf 2} \chi^r_{\bf 3}+2 \chi^l_{\bf 2}+2 \chi^r_{\bf 2}-2 \chi^r_{\bf 4}+\chi^r_{\bf 6})\chi^{SU(3)}_{\bf \bar{3}}
\Big) + \mathcal{O}(q^2) \\
\frac{\mathcal{W}_{\mathbf{8}}^{U(3)}}{\mathcal{W}_{\mathbf{1}}^{U(3)}} &= \chi_{\mathbf{8}}^{SU(3)} + q \cdot \frac{t^6 \cdot  (\chi^f_{\bf 2}-\chi ^l_{\bf 2})}{\prod_{i\neq j}^3 (1-t^2w_iw_j^{-1})} \Big(
(\chi^r_{\bf 2}-\chi^f_{\bf 2})(\chi^{SU(3)}_{\bf 35} + \chi^{SU(3)}_{\bf \overline{35}})
+ (2 \chi^f_{\bf 2} \chi^r_{\bf 3}-7 \chi^f_{\bf 2}+2 \chi^l_{\bf 2}-2 \chi^r_{\bf 4}) \chi^{SU(3)}_{\bf 27}\nonumber\\&
+ (4 \chi^r_{\bf 2}-\chi^l_{\bf 2} \chi^f_{\bf 2}  \chi^r_{\bf 2}+4 \chi^f_{\bf 2} \chi^r_{\bf 3}+\chi^f_{\bf 3} \chi^r_{\bf 2}-4 \chi^f_{\bf 2}-\chi^l_{\bf 2} \chi^r_{\bf 3}+\chi^l_{\bf 2})
(\chi^{SU(3)}_{\bf 10} + \chi^{SU(3)}_{\bf \overline{10}}) + (2 \chi^f_{\bf 2} \chi^l_{\bf 2} \chi^r_{\bf 4}-\chi^l_{\bf 2}\chi^f_{\bf 2}\chi^r_{\bf 2}
\nonumber\\&
+\chi^f_{\bf 3} \chi^l_{\bf 2}+5 \chi^f_{\bf 2} \chi^r_{\bf 3}-3 \chi^f_{\bf 2} \chi^r_{\bf 5}+\chi^f_{\bf 3} \chi^r_{\bf 2}-2 \chi^f_{\bf 3} \chi^r_{\bf 4}-12 \chi^f_{\bf 2}-2 \chi^l_{\bf 2} \chi^r_{\bf 3}+3 \chi^l_{\bf 2}+3 \chi^r_{\bf 2}-6 \chi^r_{\bf 4}) \chi^{SU(3)}_{\bf 8}
-\chi^l_{\bf 2}\chi^f_{\bf 2} \chi^r_{\bf 2}\nonumber\\&
-\chi^f_{\bf 2} \chi^l_{\bf 2} \chi^r_{\bf 6}-\chi^f_{\bf 3} \chi^l_{\bf 2} \chi^r_{\bf 3}-\chi^f_{\bf 3} \chi^l_{\bf 2} \chi^r_{\bf 5}+4 \chi^f_{\bf 2} \chi^r_{\bf 3}+2 \chi^f_{\bf 2} \chi^r_{\bf 5}+2 \chi^f_{\bf 2} \chi^r_{\bf 7}+\chi^f_{\bf 3} \chi^r_{\bf 2}+\chi^f_{\bf 3} \chi^r_{\bf 6}-3 \chi^f_{\bf 2}+\chi^l_{\bf 2} \chi^r_{\bf 3}+\chi^l_{\bf 2} \chi^r_{\bf 5}+2 \chi^l_{\bf 2}\nonumber\\&
+3 \chi^r_{\bf 2}+2 \chi^r_{\bf 4}+5 \chi^r_{\bf 6}+2 \chi^r_{\bf 8}
 \Big) + \mathcal{O}(q^2) 
\end{align}
These results can be combined into the line defect partition function for $U(2)$ and $U(3)$ gauge theories.
%by \eqref{eq:part-1d5d}. 

Finally, the complete 5d BPS spectrum with a line defect can be obtained by multiplying the 5d/1d partition function of \eqref{eq:part-1d5d}  with the perturbative W-bosons' partition function $Z_{\rm pert}$. This perturbative piece can be computed using the equivariant index theorem \cite{Shadchin:2004yx} or equivalently by counting the BPS letter operators, and is given as
\begin{align}
	Z_{\rm pert}^{U(N)} = \text{PE}\left[\frac{\text{sh}(M\pm \epsilon_+)}{\text{sh}(\epsilon_+\pm \epsilon_-)}  \cdot \Big(\chi_{\mathbf{adj}}^{SU(N)} + 1\Big)^+ \right]  = \text{PE}\left[\frac{t(\mu+\mu^{-1}-t-t^{-1})}{(1-tu)(1-tu^{-1})}  \cdot \Big(\chi_{\mathbf{adj}}^{SU(N)} + 1\Big)^+ \right].
\end{align}
The superscript `$+$' means those non-BPS states corresponding to non-positive roots are all discarded.

\section{Wilson surfaces in 6d $\mathcal{N}=(2,0)$ SCFTs}
\label{sec:6d}

\subsection{Gauge theories on self-dual strings}
\label{subsec:2dglsm}
\begin{figure}[b]
        \centering
        \includegraphics[]{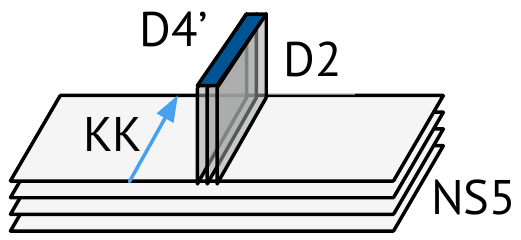}
        \caption{The brane configuration describing codimension-4 Wilson surfaces in 6d $(2,0)$ theories.  }
        \label{fig:WilsonSurfaceBrane}
\end{figure}

In order to engineer Wilson surfaces in 6d $(2,0)$ theories, we start from the D0-D4-D4$'$ brane configuration (Table~\ref{tbl:brane-d0d4d4'}) that engineers the 5d $\mathcal{N}=1^*$ $U(N)$ gauge theory with the $\frac{1}{2}$-BPS line defects
%We first T-dualize along the $x^5$ direction, then take the S-duality transformation which exchanges the $x^5$ circle and the M-theory circle. The dual system is given by a type IIA configuration composed of NS5-D6-D4$'$-D2 branes (Table~\ref{tbl:brane-d2ns5d6d4'}). 
and T-dualize along the $x^5$ direction, then take S-duality followed by another T-duality along $x^5$. The dual system is given by a type IIA configuration composed of NS5-D6-D4$'$-D2 branes (See Table~\ref{tbl:brane-d2ns5d6d4'} as well as Figure~\ref{fig:WilsonSurfaceBrane}). 
The brane system preserves the $SO(1,1)_{01} \times SO(4)_{2345} \times SO(3)_{678}$ symmetry. It is often convenient to write $SO(4)_{2345} \simeq SU(2)_{1L} \times SU(2)_{1R}$ and $SO(3)_{678} \simeq SU(2)_{2R}$. We denote the doublet indices of $SU(2)_{1L}$, $SU(2)_{1R}$, $SU(2)_{2R}$ by $\alpha, \dot{\alpha}, A$, respectively. 
The 32 supercharges of IIA string theory can be written as $Q^{\alpha A}_{\pm \pm}$ and $Q^{\dot{\alpha}A}_{\pm \pm}$, where the first/second subscripts are eigenvalues of $\Gamma^{01}$ and $\Gamma^9$. The independent SUSY projectors imposed by NS5-, D6-, D2-branes are $\Gamma^{01}$, $\Gamma^{2345}$, $\Gamma^9$. The presence of D4$'$-branes does not yield an additional SUSY projector. The surviving supercharges are $Q^{\dot{\alpha}A}_{-+}$, satisfying 2d $\mathcal{N}=(0,4)$ supersymmetry algebra.

\begin{table}[htbp]
	\centering
	\begin{tabular}{c|cccccccccc}
		& 0 & 1 & 2 & 3 & 4 & 5 & 6 & 7 & 8 & 9\\ \hline
		$N$ NS5 & $\bullet$ & $\bullet$ & $\bullet$ & $\bullet$ & $\bullet$ & $\bullet$ &  &  &  & \\
	    $N'$ D4$'$ & $\bullet$ & $\bullet$ & & & &   & $\bullet$ & $\bullet$ & $\bullet$ & \\
		1 D6 & $\bullet$ & $\bullet$ & $\bullet$ & $\bullet$ & $\bullet$ & $\bullet$ &  &  &  & $\bullet$ \\
		D2 & $\bullet$ &$\bullet$ & & & &  &  &  &  & $\bullet$
	\end{tabular}
	\caption{Brane configuration for 6d $\mathcal{N}=(2,0)$ SCFTs with Wilson surface defects.}
	\label{tbl:brane-d2ns5d6d4'}
\end{table}

The parallel stack of $N$ NS5-branes engineers 6d $\mathcal{N}=(2,0)$ SCFT of $A_{N-1}$ type. We separate the NS5-branes from one another along the $x^9$ direction. The distances between adjacent NS5-branes are parametrized by the VEV of the tensor multiplet scalar $\Phi_9$. It spans the tensor branch of $(2,0)$ SCFT where the conformal symmetry is spontaneously broken. The D6-brane in Table~\ref{tbl:brane-d2ns5d6d4'} is obtained from dualizing the Taub-NUT space of Table~\ref{tbl:brane-d0d4d4'}.
%, providing the mass deformation of 5d $\mathcal{N}=2$ gauge theories. 
The BPS strings of 6d SCFTs are electric/magnetic sources of the tensor multiplets. They are realized by D2-branes suspended between an adjacent pair of NS5-branes. In the tensor branch, they acquire a non-zero tension which is proportional to the VEV $\langle \Phi_9 \rangle \neq 0$. The worldvolume gauge theory of D2-branes \cite{Haghighat:2013gba,Haghighat:2013tka}
turns out to be useful for studying the protected quantities of 6d BPS strings, such as the chiral anomaly, central charge and the elliptic genus. 

The D4$'$-branes of Table~\ref{tbl:brane-d2ns5d6d4'} are obtained from the D4$'$-branes of Table~\ref{tbl:brane-d0d4d4'}. These are located at the origin $x^2 = \cdots = x^5 =0$ of the $\mathbf{R}^4$ plane. 
%\cmt{We will explain later that the numbers of D4$'$-branes in the $x^9$ intervals between adjacent NS5-branes determine the representation of the Wilson surface operators.} 
The fundamental string stretched between the D4-D4$'$ system of section \ref{sec:5dWilson} now becomes a D2-brane stretched between the D4$'$ branes and the NS5 branes. From the point of view of the world-volume theory of NS5 branes, this corresponds to the creation of heavy probe strings whose presence leads to the Wilson surfaces in the 6d (2,0) theory. The representation carried by these Wilson surfaces can be understood in the following way: To begin with note that there can at most be a single D2-brane stretched between an NS5-brane and a D4$'$-brane. This is because such a D2-brane is fermionic in nature and therefore this system is governed by the S-rule of Hanany and Witten \cite{Hanany:1996ie}. Now, a single D2-brane ending on the stack of NS5-branes creates a Wilson surface in the fundamental representation. When there are $m$ D2-branes stretched from a single D4$'$-brane to the stack of NS5 branes, the fermionic nature of D2-branes implies that the Wilson surface is in the rank-$m$ anti-symmetric representation.  As a consequence of S-rule, $m$ is constrained to be $m \leq N$. 

\begin{figure}[t]
	\centering
    \begin{subfigure}[b]{1\textwidth}
    \centering
	\includegraphics[width=5.0in]{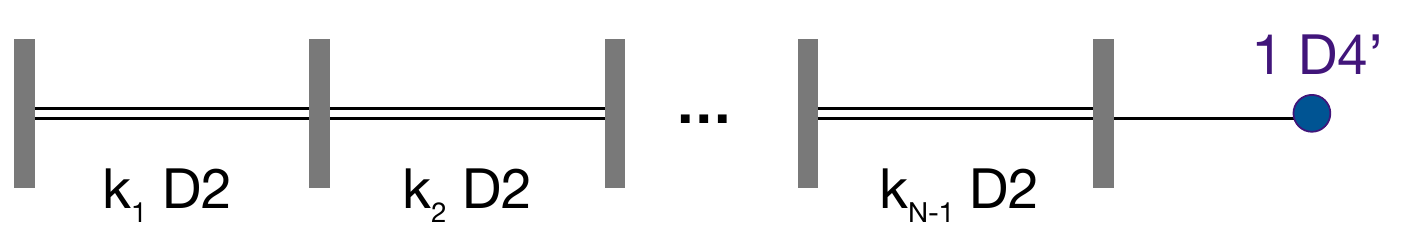}
    \caption{The brane configuration engineering the fundamental Wilson surface of the 6d $(2,0)$ theory of type $A_{N-1}$. The vertical lines in the picture denote NS5-branes. }
    \label{fig:FundamentalWilsonSurfaceA}
    \end{subfigure}
    ~
    \begin{subfigure}[b]{1\textwidth}
    \centering
	\includegraphics[width=4.0in]{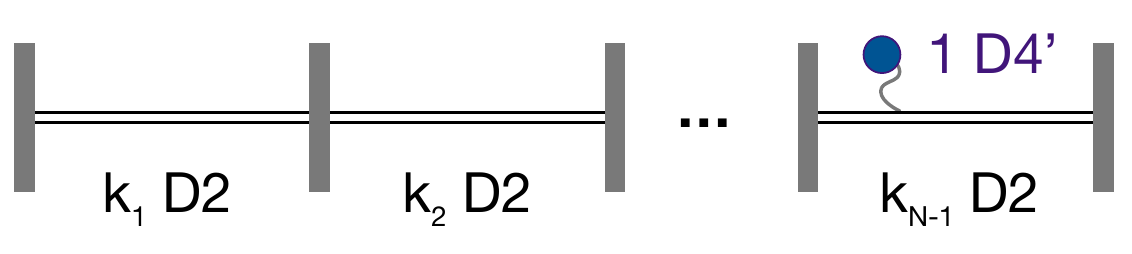}
    \caption{The brane configuration obtained from application of Hanany-Witten transition to the set-up of figure \ref{fig:FundamentalWilsonSurfaceA}. The vertical lines in the picture denote NS5-branes. \newline }
    \label{fig:FundamentalWilsonSurfaceB}
    \end{subfigure}
    ~
    \begin{subfigure}[b]{1\textwidth}
    \centering
	\includegraphics[width=4.0in]{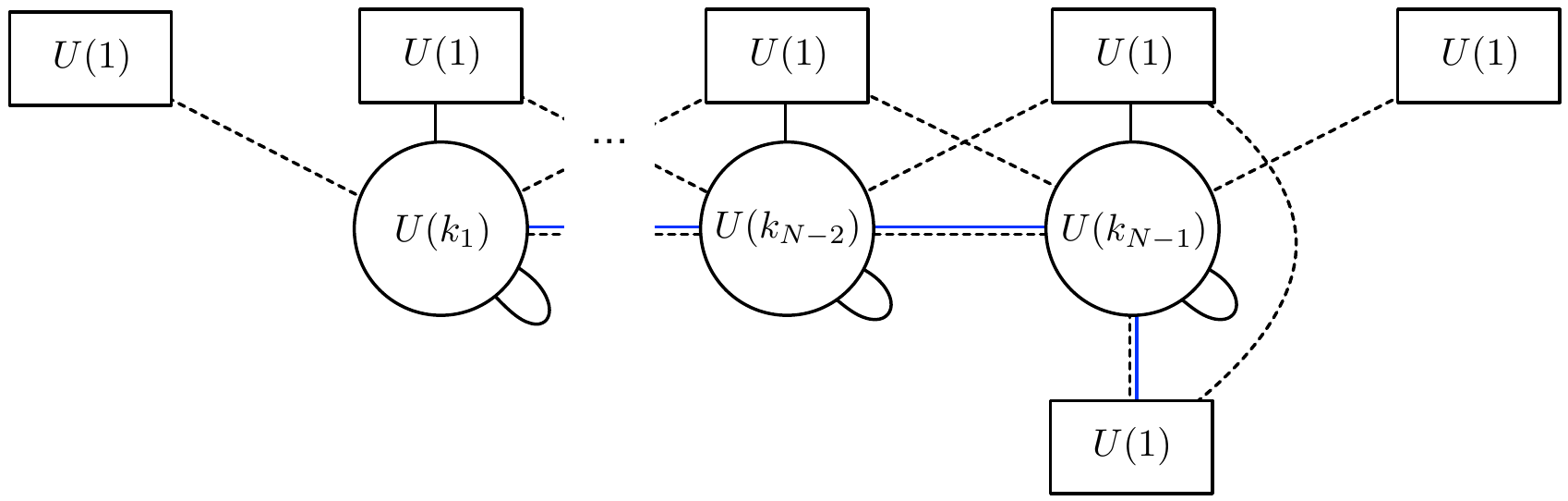}
    \caption{The 2d $(0,4)$ quiver gauge theory obtained from the brane configuration of figure \ref{fig:FundamentalWilsonSurfaceB}. Here solid black lines denote a hypermultiplet, solid blue lines denote a twisted hyper-multiplet, while the dashed lines denote a Fermi multiplet.  } . 
    \label{fig:FundamentalWilsonSurfaceC}
    \end{subfigure}
    \caption{The brane configuration and 2d theory describing the fundamental Wilson surface defect. Table~\ref{tbl:brane-d2ns5d6d4'} also has a D6-brane which has not been shown here.  }
	\label{fig:FundamentalWilsonSurface}
    \end{figure}

The above description can now be used to obtain the 2d quivers describing the effect of Wilson-surface insertion into the elliptic genera of 6d self-dual strings. Once again, let us begin by describing the fundamental Wilson surface. It will be convenient to label the NS5 branes from 1 to $N$ with the order of numbering being from left to right. We then introduce a single D4$'$-brane to the right of this stack and stretch a D2-brane connecting the $N$-th NS5-brane and the D4$'$-brane. Let us also suspend $k_i$ D2-branes between $i$'th and $(i+1)$'th NS5-branes. We show this brane configuration in Figure \ref{fig:FundamentalWilsonSurfaceA}. 
The 2d quiver obtained from this configuration is expected to engineer the insertion of fundamental Wilson surface into the elliptic genera. As we will momentarily explain, this is a $\prod_{i=1}^{N-1} U(k_i)$ quiver gauge theory. In the absence of the D4$'$-brane, this quiver was first described in \cite{Haghighat:2013gba,Haghighat:2013tka}. The extra matter introduced into the quiver as a result of including the D4$'$-branes can be seen by first using Hanany-Witten transition to go to the frame where the D4$'$-brane lies between the $(N-1)$'th and the $N$'th NS5-brane (see Figure \ref{fig:FundamentalWilsonSurfaceB}).  
The D2-brane suspended between the D4$'$-brane and NS5-brane is annihilated in the process. Open strings stretched between the D4$'$-brane and the stack of $k_{N-1}$ D2-branes (suspended between the $(N-1)$'th and $N$'th NS5 branes) will now give rise to a $U(k_{N-1})$-fundamental $(0,4)$ twisted hyper and a $U(k_{N-1})$-fundamental Fermi multiplet transforming as a $SU(2)_{1L}$ doublet. Additionally, there are gauge singlet Fermi multiplets coming from the excitations of the open string stretched between the D4$'$-brane and the D6-brane.   The resulting quiver is shown in Figure \ref{fig:FundamentalWilsonSurfaceC}. 
%The insertion of a fundamental Wilson surface into the elliptic genera of 6d strings is described by including these extra superfields in the quiver of \cite{Haghighat:2013gba,Haghighat:2013tka}.    

\begin{figure}[t]
	\centering
    \begin{subfigure}[b]{1\textwidth}
    \centering
	\includegraphics[width=5.0in]{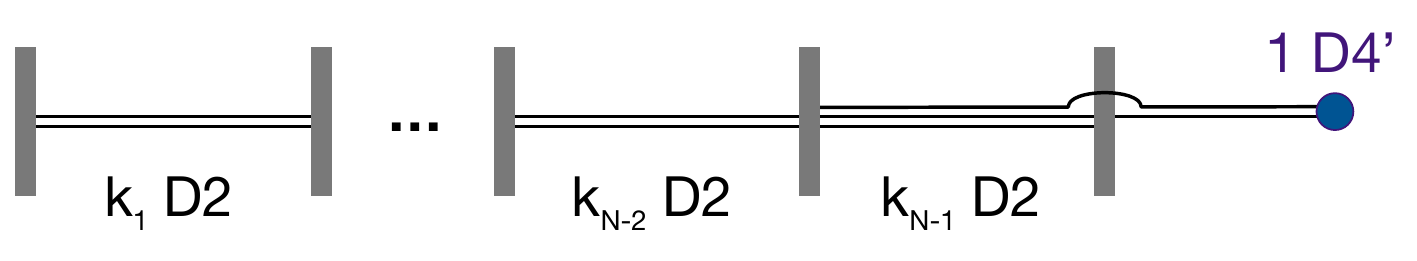}
    \caption{The brane configuration engineering the rank-2 antisymmetric  Wilson surface of the 6d $(2,0)$ theory of type $A_{N-1}$. The vertical lines in the picture denote NS5-branes. }
    \label{fig:AntiSymWilsonSurfaceA}
    \end{subfigure}
    ~
    \begin{subfigure}[b]{1\textwidth}
    \centering
	\includegraphics[width=4.0in]{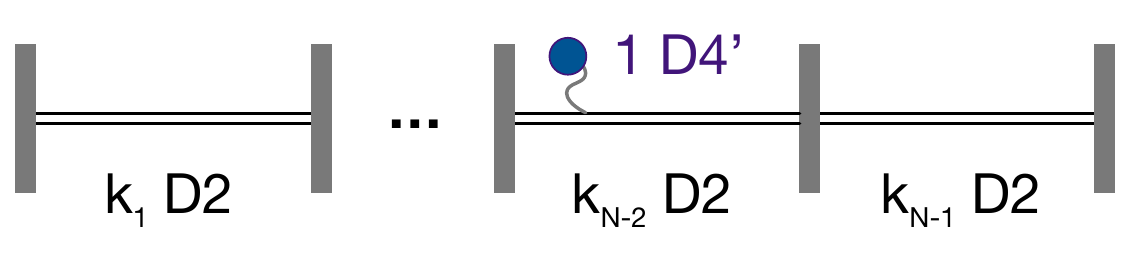}
    \caption{The brane configuration obtained from application of Hanany-Witten transition to the set-up of figure \ref{fig:AntiSymWilsonSurfaceA}. The vertical lines in the picture denote NS5-branes. \newline . }
    \label{fig:AntiSymWilsonSurfaceB}
    \end{subfigure}
    ~
    \begin{subfigure}[b]{1\textwidth}
    \centering
	\includegraphics[width=4.0in]{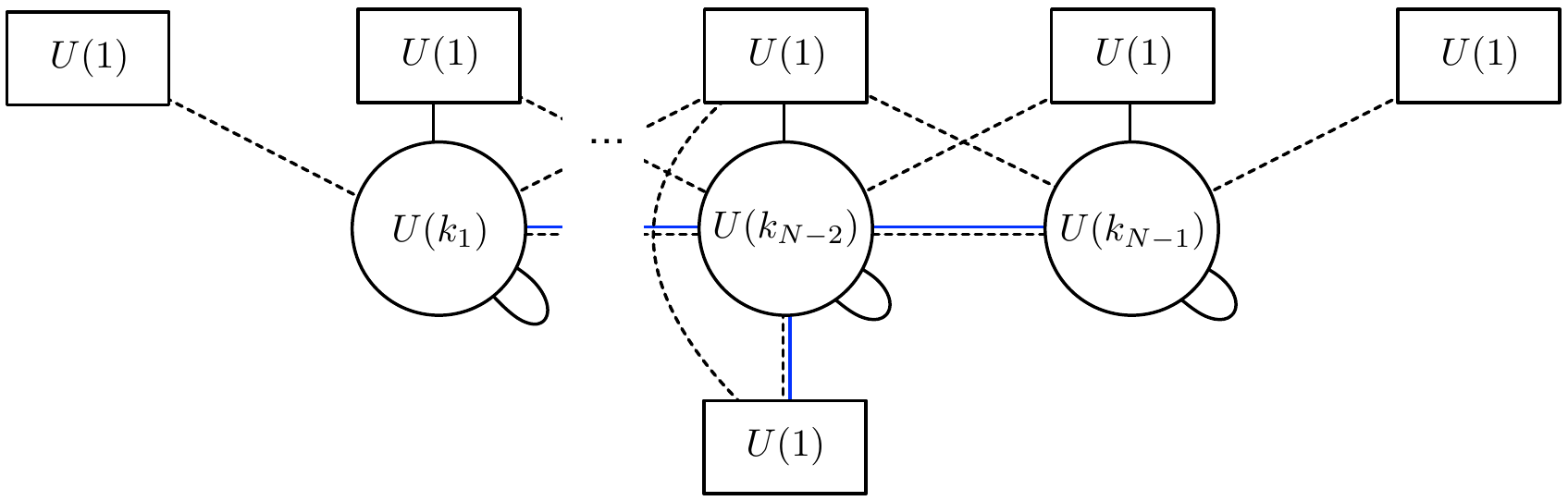}
    \caption{The 2d $(0,4)$ quiver gauge theory obtained from the brane configuration of figure \ref{fig:AntiSymWilsonSurfaceB}. Here solid black lines denote a hypermultiplet, solid blue lines denote a twisted hyper-multiplet, while the dashed lines denote a Fermi multiplet.  }  
    \label{fig:AntiSymWilsonSurfaceC}
    \end{subfigure}
    \caption{The brane configuration and 2d theory describing the rank-2 antisymmetric Wilson surface defect. Table~\ref{tbl:brane-d2ns5d6d4'} also has a D6-brane which has not been shown here.  }
	\label{fig:AntiSymWilsonSurface}
    \end{figure}

Similarly, we can engineer the Wilson surface in the rank-$m$ anti-symmetric irreducible representation by starting with a D4$'$-brane lying to the right of the NS5-brane stack and stretching $m$ D2-branes from the D4$'$ to the NS5-brane stack such that the $i$'th D2-brane ends on the $(N+1-i)$'th NS5 brane, for all $1\leq i \leq m$. This set-up can be dualised by the Hanany-Witten transition, to a frame where the D4$'$-brane lies between the $(N-m)$'th and $(N+1-m)$'th NS5-brane with no D2-branes suspended between the D4$'$ and NS5-branes. Open strings stretched between the D4$'$ and the $k_{N-m}$ D2-branes (suspended between the $(N-m)$'th and $(N+1-m)$'th NS5 branes) will then give rise to a $U(k_{N-m})$-fundamental twisted hyper and a $U(k_{N-m})$-fundamental Fermi multiplet  in $SU(2)_{1L}$ doublet. Also, the open strings stretched between the D4$'$-brane and the D6-brane give rise to a Fermi multiplet transforming as the bifundamental of $U(1)_{N-m} \times U(1)$, where the $U(1)_{N-m}$ factor comes from the $U(1)$ symmetry associated to the D6 brane segment between the $(N-m)$'th and the $(N+1-m)$'th NS5-brane and the other $U(1)$ factor comes from the symmetry associated to the D4$'$-brane. The inclusion of these fields in the 2d quiver will therefore capture the effect of a rank-$m$ antisymmetric Wilson surface. We show the relevant brane configuration and 2d quiver for the case of a rank-2 antisymmetric Wilson surface in  Figure \ref{fig:AntiSymWilsonSurface}.  
    
\begin{figure}[h]
	\centering
    \begin{subfigure}[b]{1\textwidth}
    \centering
	\includegraphics[width=4.0in]{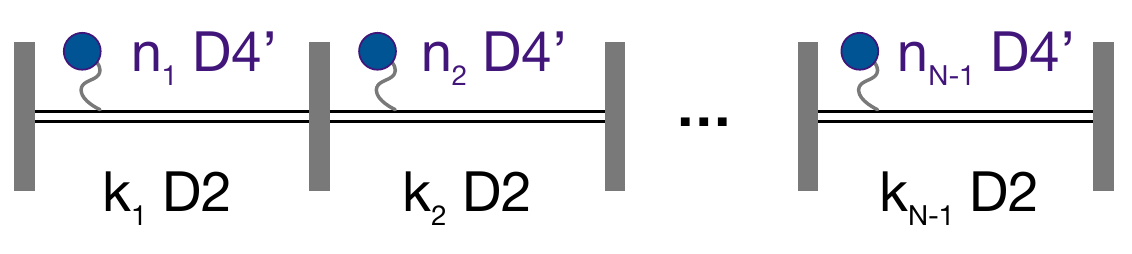}
    \caption{The brane configuration for the Wilson surface in the representation $\mathbf{R} = \otimes_{i=1}^{N-1}  \left(\bigwedge^{N-i}\right)^{n_i}$. The vertical lines in the picture denote NS5-branes. \newline . }
    \label{fig:GenWilsonSurfaceA}
    \end{subfigure}
    ~
    \begin{subfigure}[b]{1\textwidth}
    \centering
	\includegraphics[width=4.0in]{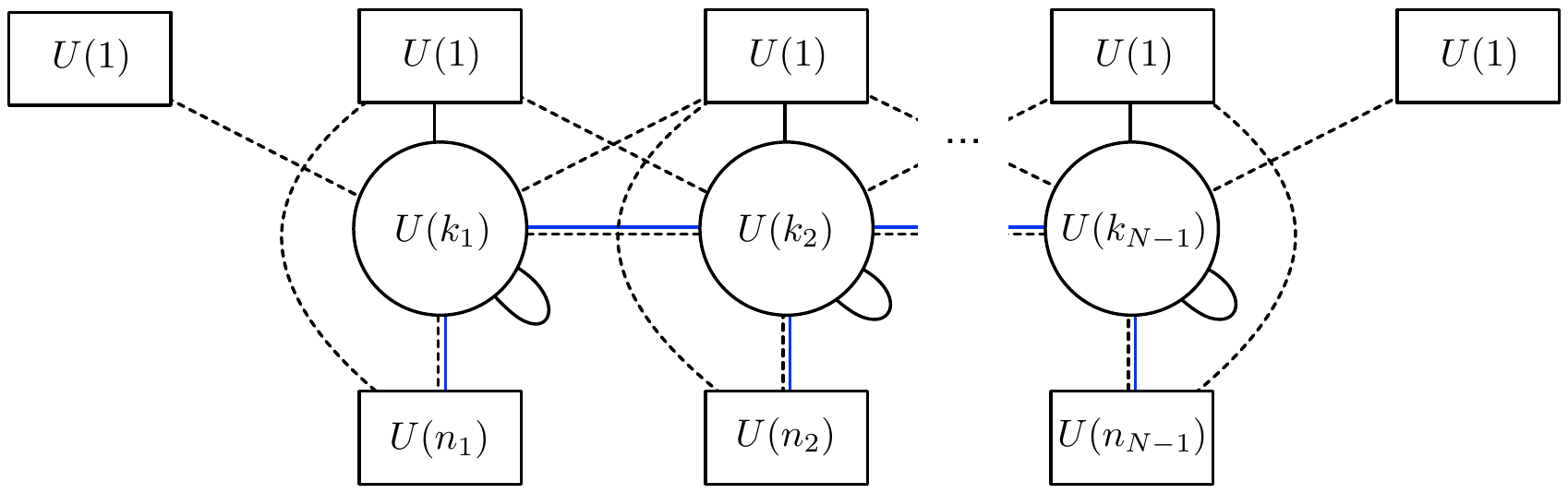}
    \caption{The 2d $(0,4)$ quiver gauge theory obtained from the brane configuration of figure \ref{fig:GenWilsonSurfaceA}. Here solid black lines denote a hypermultiplet, solid blue lines denote a twisted hyper-multiplet, while the dashed lines denote a Fermi multiplet.  } 
    \label{fig:GenWilsonSurfaceB}
    \end{subfigure}
    \caption{The brane configuration and 2d theory describing a Wilson surface defect in the representation $\mathbf{R} = \otimes_{i=1}^{N-1}  \left(\bigwedge^{N-i}\right)^{n_i}$. Table~\ref{tbl:brane-d2ns5d6d4'} also has a D6-brane which has not been shown here.  }
	\label{fig:GenWilsonSurface}
    \end{figure}

In the general case of $N' > 1$, we can consider brane configurations with $n_i$ D4$'$-branes between the $i$'th and the $(i+1)$'th NS5 brane, such that $\sum_{i = 1}^{N-1} n_i = N'$ (see Figure \ref{fig:GenWilsonSurfaceA}); from the above discussion it is clear that such configuration will correspond to inserting a Wilson surface in the representation $\mathbf{R} = \otimes_{i=1}^{N-1}  \left(\bigwedge^{N-i}\right)^{n_i}$. The resulting 2d quiver is shown in Figure \ref{fig:GenWilsonSurfaceB}.

%From the above discussion it is clear that the brane configuration with $n_i$ D4$'$-branes between the $i$'th and the $(i+1)$'th NS5 brane, for all $1 \leq i \leq N$, will correspond to inserting a Wilson surface in the representation given by $\mathbf{R} = \otimes_{i=1}^{N}  \left(\Lambda^{N+1-i}\right)^{n_i}$.

%For the moment, we introduce the 2d gauge theories of D2-branes suspended between NS5-branes with insertion of D4$'$-branes. 
Let us now give a more detailed description of the 2d gauge theories obtained from the world-volume of D2-branes suspended between NS5-branes along with the insertion of D4$'$-branes. 
We consider a stack of $k_i$ D2-branes and $n_i$ D4$'$-branes, located in the $x^9$ interval between $i$'th and $(i+1)$'th NS5-branes, for all $i=1,\, \cdots,\, N-1$. We draw the given configuration in Figure~\ref{fig:GenWilsonSurfaceA}. The chain of $(k_1, k_2, \cdots, k_{N-1})$ D2-branes engineers a 2d linear quiver theory with $\prod_{i=1}^{N-1} U(k_i)$ gauge symmetry. It inherits $SU(2)_{1L} \times SU(2)_{1R} \times SU(2)_{2R}$ global symmetry of the brane system, where $SU(2)_{1R} \times SU(2)_{2R}$ is the R-symmetry of 2d $\mathcal{N}=(0,4)$ supersymmetry.
At a classical level, it also captures $\prod_{i=1}^{N-1} U(n_i) \times \prod_{j=0}^{N} U(1)_j$ flavor symmetry of $(n_1, n_2, \cdots, n_{N-1})$ D4$'$-branes and chain of $(N+1)$ D6-brane segments. 
Massless excitations coming from open strings ending on D2-branes induce the field content of Table~\ref{tbl:2d-field}, in which 
the $U(1)_i$ charge of an individual field is denoted by $\mathcal{Q}_i$. Only a subset of global $U(1)$'s will be consistent at a quantum level, yielding no mixed anomaly with gauge symmetry.
If one uses $T_i$, $F_i$, $G_i$ to denote the Abelian generators of $U(1) \subset U(k_i)$, $U(n_i)$, $U(1)_i$, respectively, then all the non-vanishing mixed anomalies between gauge and global symmetries are 
\begin{align}
\text{Tr}(\gamma_3 \,T_i G_{i-1}) = -1, \quad\text{Tr}(\gamma_3\, T_i G_{i})=2, \quad \text{Tr}(\gamma_3\, T_i G_{i+1}) =-1\qquad\text{ for all }1\leq i \leq N-1.
\end{align}
This implies that only two $U(1)$'s, out of $\prod_{j=0}^{N} U(1)_j$ global symmetry, will remain at a quantum level.
These two $U(1)$'s are the ones generated by $\sum_{j=0}^{N}G_j$ and $\sum_{j=0}^{N} (j-1)G_j$. We also notice that the sum of all $U(1)$ generators $\sum_{j=0}^{N} G_j + \sum_{i=1}^{N-1} (T_i + F_i)$ trivially acts on the fields, therefore $\sum_{j=0}^{N} G_j \simeq -\sum_{i=1}^{N-1} (T_i + F_i)$ effectively.

\begin{table}[tbp]
\centering
\begin{subtable}[t]{.55\textwidth}
\vspace{0pt}
\centering
\begin{tabular}{c|c|ccc}
Type & Fields & $U(k_i)$ & $U(n_i)$ & $(\mathcal{Q}_{i-1}, \mathcal{Q}_i, \mathcal{Q}_{i+1})$ \\\hline
vector & $A_\mu^{(i)}, \zeta_{\dot{\alpha}}^{{(i)}A}$ & \textbf{adj} & $\mathbf{1}$ & $(0,0,0)$\\
hyper & $a_{\alpha\dot{\alpha}}^{(i)}, \lambda_{\alpha}^{{(i)}A}$ & \textbf{adj} & $\mathbf{1}$ & $(0,0,0)$\\
twisted & $\Phi_{A}^{(i)}, \Psi_{\dot{\alpha}}^{(i)}$ & $\mathbf{k}_i$ & $\overline{\mathbf{n}}_i$ & $(0,0,0)$\\
Fermi & $\Psi_{\alpha}^{(i)}$ &  $\mathbf{k}_i$ & $\overline{\mathbf{n}}_i$ & $(0,0,0)$\\
hyper & $q_{\dot{\alpha}}^{(i)}, \psi^{(i)A}$ &  $\mathbf{k}_i$ & $\mathbf{1}$  & $(0,+1,0)$\\
Fermi & $\chi^{(i)}$ &  $\mathbf{k}_i$ & $\mathbf{1}$  & $(0,0,-1)$\\
Fermi & $\overline{\chi}^{(i)}$ &  $\overline{\mathbf{k}}_i$ & $\mathbf{1}$  & $(+1,0,0)$
\end{tabular}
\subcaption{in the $i$'th gauge node ($i \leq N-1$)}
\label{tbl:2d-field-node}
\end{subtable}
\hfill
\begin{subtable}[t]{.43\textwidth}
\vspace{0pt}
\centering
\begin{tabular}{c|c|cc}
Type & Fields & $U(k_i)$ & $U(k_{i+1})$ \\\hline
twisted & $\varphi_{A}^{(i)}, \xi_{\dot{\alpha}}^{(i)}$ & $\mathbf{k}_i$ & $\overline{\mathbf{k}}_{i+1}$\\
Fermi & $\xi_{\alpha}^{(i)}$ &  $\mathbf{k}_i$ & $\overline{\mathbf{k}}_{i+1}$\\
\end{tabular}
\subcaption{between $i$'th and $(i+1)$'th nodes ($i \leq N-2$)}
\label{subtbl:2d-field-bif}
\vspace{1.68\baselineskip}
\centering
\begin{tabular}{c|c|cc}
Type & Fields & $U(n_i)$ & $\mathcal{Q}_i$ \\\hline
Fermi & $\rho^{(i)}$ &  $\mathbf{n}_i$ & $-1$\\
\end{tabular}
\subcaption{between $i$'th D6- and D4-branes ($i \leq N-1$)}
\label{subtbl:2d-field-d6d4}
\end{subtable}
\caption{$\mathcal{N}=(0,4)$ multiplets in the quiver gauge theory. `twisted' means a twisted hypermultiplet.}
\label{tbl:2d-field}
\end{table}

One can easily construct the Lagrangian of the quiver gauge theory. We will use the $\mathcal{N}=(0,2)$ SUSY formalism \cite{Witten:1993yc} and write the holomorphic potentials $E_\eta$ and $J_\eta$, for each $(0,2)$ Fermi multiplet $\eta$, in a way which ensures the $\mathcal{N}=(0,4)$ SUSY enhancement \cite{Tong:2014yna}. 
First, we decompose the $(0,4)$ vector and (twisted) hypermultiplets into the $(0,2)$ chiral and Fermi multiplets as follows:
\begin{align}
\text{ vector } (A_\mu^{(i)}, \zeta_{\dot{\alpha}}^{{(i)}A}) &\longrightarrow\ \text{ vector } V^{(i)}\ (A_\mu^{(i)}, \zeta^{(i)1}_{\dot{1}}) \ \oplus\text{ Fermi }\Lambda^{(i)}\ (\zeta^{(i)1}_{\dot{2}}) \nonumber\\
\text{ hyper } (a_{\alpha\dot{\alpha}}^{(i)}, \lambda_{\alpha}^{{(i)}A})& \longrightarrow \ \text{ chiral } B^{(i)}\ (a_{1\dot{1}}^{(i)}, \lambda_{1}^{(i)2})\  \oplus \text{ chiral }\tilde{B}^{(i)\dagger}\ (a_{1\dot{2}}^{(i)}, \lambda_{1}^{(i)1}) \nonumber\\
\text{ hyper } (q_{\dot{\alpha}}^{(i)}, \psi^{(i)A})& \longrightarrow \ \text{ chiral } q^{(i)}\ (q_{\dot{1}}^{(i)}, \psi^{(i)2})\  \oplus \text{ chiral }\tilde{q}^{(i)\dagger}\ (q_{\dot{2}}^{(i)}, \psi^{(i)1})\\
\text{ twisted hyper } (\Phi_{A}^{(i)}, \Psi_{\dot{\alpha}}^{(i)})& \longrightarrow \ \text{ chiral } \Phi^{(i)}\ (\Phi_{1}^{(i)}, \Psi_{\dot{1}}^{(i)})\  \oplus \text{ chiral }\tilde{\Phi}^{(i)\dagger}\ (\Phi_{2}^{(i)}, \Psi_{\dot{2}}^{(i)})\nonumber\\
\text{ twisted hyper } (\varphi_{A}^{(i)}, \xi_{\dot{\alpha}}^{(i)})& \longrightarrow \ \text{ chiral } \varphi^{(i)}\ (\varphi_{1}^{(i)}, \xi_{\dot{1}}^{(i)})\  \oplus \text{ chiral }\tilde{\varphi}^{(i)\dagger}\ (\varphi_{2}^{(i)}, \xi_{\dot{2}}^{(i)})\nonumber
\end{align}
On the other hand, all $(0,4)$ Fermi multiplets $\Psi_{\alpha}^{(i)}$, $\chi^{(i)}$, $\overline{\chi}^{(i)}$, $\xi_{\alpha}^{(i)}$, $\rho^{(i)}$ are also $(0,2)$ Fermi multiplets. 
Second, we introduce $E_\Lambda^{(i)}$ and $J_\Lambda^{(i)}$ for Fermi multiplets $\Lambda^{(i)}$ as functions of  $(0,2)$ chiral superfields. 
We follow \cite{Tong:2014yna} which introduces the chiral multiplets from (twisted) hypermultiplets to $J_\Lambda^{(i)}$ ($E_\Lambda^{(i)}$), i.e.,
\begin{align}
J_\Lambda^{(i)} = q^{(i)}\tilde{q}^{(i)} + [B^{(i)}, \tilde{B}^{(i)}], \qquad E_\Lambda^{(i)} = \varphi^{(i)}\tilde{\varphi}^{(i)} - \tilde{\varphi}^{(i-1)}\varphi^{(i-1)} + \Phi^{(i)} \tilde{\Phi}^{(i)}
\end{align}
with $\varphi^{(0)}$, $\tilde{\varphi}^{(0)}$, $\varphi^{(N)}$, $\tilde{\varphi}^{(N)}$ understood as being null. Third, the above $J_\Lambda^{(i)}$ and $E_\Lambda^{(i)}$ must be accompanied by other holomorphic potentials to satisfy the $(0,2)$ SUSY preserving condition $\sum_\eta \text{Tr} \,(J_\eta \cdot E_\eta) = 0$, where the sum is taken over all Fermi multiplets in the theory.
Allowed are the following potentials:
\begin{gather}
E_{\chi}^{(i)} = \varphi^{(i)}q^{(i+1)},\quad J_{\chi}^{(i)} = \tilde{q}^{(i+1)}\tilde{\varphi}^{(i)}, \quad 
E_{\overline{\chi}}^{(i)} = \tilde{q}^{(i-1)}\varphi^{(i-1)}, \quad J_{\overline{\chi}}^{(i)} = -\tilde{\varphi}^{(i-1)}q^{(i-1)},\nonumber\\
E_{\Psi_1}^{(i)} = B^{(i)}\Phi^{(i)}, \quad J_{\Psi_1}^{(i)} = \tilde{\Phi}^{(i)}\tilde{B}^{(i)},\quad 
E_{\Psi_2}^{(i)} = -\tilde{B}^{(i)}\Phi^{(i)}, \quad J_{\Psi_2}^{(i)} = \tilde{\Phi}^{(i)}B^{(i)},\\
E_{\xi_1}^{(i)} = \varphi^{(i)}B^{(i+1)} - B^{(i)}\varphi^{(i)}, \quad E_{\xi_2}^{(i)} = \tilde{B}^{(i)}\varphi^{(i)}-\varphi^{(i)}\tilde{B}^{(i+1)},\quad E_{\rho}^{(i)} = \tilde{\Phi}^{(i)}q^{(i)}, \nonumber\\
J_{\xi_1}^{(i)} = \tilde{B}^{(i+1)}\tilde{\varphi}^{(i)}- \tilde{\varphi}^{(i)}\tilde{B}^{(i)},
\quad J_{\xi_2}^{(i)} = B^{(i+1)}\tilde{\varphi}^{(i)}- \tilde{\varphi}^{(i)}B^{(i)},
\quad J_{\rho}^{(i)} = -\tilde{q}^{(i)}\Phi^{(i)}.\nonumber
\end{gather}
With the following D-term potential $D^{(i)}$,
\begin{align}
D^{(i)} &= q^{(i)}q^{(i)\dagger}-\tilde{q}^{(i)\dagger}\tilde{q}^{(i)} + [B^{(i)},B^{(i)\dagger}] - [\tilde{B}^{(i)\dagger},\tilde{B}^{(i)}]\\&
+ \Phi^{(i)}\Phi^{(i)\dagger}-\tilde{\Phi}^{(i)\dagger}\tilde{\Phi}^{(i)} - \varphi^{(i-1)\dagger}\varphi^{(i-1)}+\tilde{\varphi}^{(i-1)}\tilde{\varphi}^{(i-1)\dagger} + \varphi^{(i)}\varphi^{(i)\dagger}-\tilde{\varphi}^{(i)\dagger}\tilde{\varphi}^{(i)},
\end{align}
the bosonic potential $V = \sum_{\eta} (|E_\eta|^2 + |J_\eta|^2) + \frac{1}{4}\sum_{i=1}^{N-1} (D^{(i)})^2 $ can be arranged into
\begin{align}
V = \sum_{i=1}^{N-1} \bigg( &\frac{1}{4} \Big((\sigma^m)^{\dot{\alpha}}{}_{\dot{\beta}}  q_{\dot{\alpha}}^{(i)}q^{(i)\dagger\dot{\beta}} + \frac{1}{2}(\sigma^m)^{\dot{\alpha}}{}_{\dot{\beta}} [a_{\alpha \dot{\alpha}}^{(i)}, a^{(i)\alpha\dot{\beta}}] \Big)^2 \nonumber\\
+& \frac{1}{4} \Big((\sigma^I)^{A}{}_{B} \varphi_{A}^{(i)} \varphi^{(i)\dagger B} - (\sigma^I)^{A}{}_{B}  \varphi^{(i-1)\dagger B} \varphi^{(i-1)}_A + (\sigma^I)^{A}{}_{B}  \Phi_{A}^{(i)}\Phi^{(i)\dagger B}\Big)^2 \\
+& \frac{1}{2}|q_{\dot{\alpha}}^{(i)} \Phi_{A}^{(i)}|^2 + \frac{1}{2}|q_{\dot{\alpha}}^{(i)}\varphi_A^{(i-1)\dagger}| + \frac{1}{2}|q_{\dot{\alpha}}^{(i)}\varphi_A^{(i)}| + \frac{1}{2} |a^{(i)}_{\alpha\dot{\alpha}}  \Phi_A^{(i)}  |^2 + \frac{1}{2} | \varphi_A^{(i)}a^{(i+1)}_{\alpha\dot{\alpha}}  - a^{(i)}_{\alpha\dot{\alpha}}\varphi_A^{(i)}  |^2  \bigg).\nonumber
\end{align}

Note that though we necessarily need the Fermi fields $\rho$ in order to write a Lagrangian with $(0,4)$ SUSY, we expect that these will decouple from the interacting system in the IR. A simple way to see this is to notice that they are gauge invariant chiral operators and hence their IR scaling dimensions can be deduced from their charge with respect to the R-symmetry of $(0,2)$ subalgebra. The R-current of $(0,2)$ subalgebra will be given by a linear combination of the two Cartans of the $SO(4)_R$-symmetry acting on the UV $(0,4)$ algebra. However, $\rho$ is a singlet of this $SO(4)_R$ symmetry and hence it will carry a zero charge with respect to the $(0,2)$ R-symmetry. Since the right moving dimension of chiral operators is proportional to their R-charge, it implies that for $\rho$, $\bar{L}_0 =0$. Their left moving dimension $L_0$ can also be fixed by using the fact that $L_0-\bar{L}_0=\frac{1}{2}$, since the difference of the left and right moving dimensions is given by the spin of the operators. Thus we find that the fermions $\rho$ must have $L_0=\frac{1}{2}$ and $\bar{L}_0=0$. This is the unitarity bound for left moving gauge invariant fermions to become free, much like in \cite{Bertolini:2014ela,Gadde:2016khg}.

\subsection{Elliptic genera with Wilson surfaces}

We consider the elliptic genera of 2d $\mathcal{N}=(0,4)$ quiver gauge theories, introduced in Section~\ref{subsec:2dglsm}. They are labeled by the ranks of the $N-1$ product gauge and flavor groups $\{k_1, k_2,\cdots,k_{N-1} \,|\, n_1, n_2,\cdots,n_{N-1}\}$, where $\{k_1, k_2,\cdots,k_{N-1}\}$ and $\{n_1, n_2,\cdots,n_{N-1}\}$ correspond to the winding numbers for different species of self-dual strings and defect strings.
 One can express them in the following trace representation:
\begin{align}
  \label{eq:elliptic-genera}
  \mathcal{I}_{\{k_1,\cdots,k_{N-1}\}}^{\{n_1,\cdots,n_{N-1}\}} = \text{Tr}_{RR}\,\left[ (-1)^F
q^{H_L} \bar{q}^{H_R}\,  t^{2(J_r - J_R)} u^{2J_l} \mu^{\sum_{j=0}^{N} (j-1)G_j}  \prod_{i=1}^{N-1} \prod_{\ell=1}^{n_i} e^{- m_{\ell}^{(i)} Q_{\ell}^{(i)}} \right].
\end{align}
The elliptic genus captures the BPS states annihilated by $Q^{1\dot{1}}$ and $Q^{2\dot{2}}$, satisfying $H_R \sim \{Q^{1\dot{1}}, Q^{2\dot{2}}\}$.
Most of the fugacity variables and conjugate charges were introduced in Section~\ref{subsec:adhm-index}. In addition, $m_{\ell}^{(i)}$ are chemical potentials for $U(n_i)$ flavor symmetry, whose Cartan generators are denoted as $Q_{\ell}^{(i)}$. All these deformations made by chemical potentials are crucial for regulating the infrared divergences.  

Evaluation of the elliptic genus of 2d $\mathcal{N}=(0,2)$ gauge theory was formulated in \cite{Gadde:2013ftv,Benini:2013nda,Benini:2013xpa} by means of supersymmetric localization. Saddle points in the path integral  are characterized by the gauge flat connections on $T^2$. We denote eigenvalues of the $U(k_i)$ gauge connection by $\phi^{(i)} = \{ \phi^{(i)}_1, \phi^{(i)}_2, \cdots, \phi^{(i)}_{k_i}\}$, and the collection of all gauge connections by $\phi = \{ \phi^{(1)}, \ldots, \phi^{(N-1)} \}$. All the Gaussian fluctuations around saddle points can be integrated out, leaving us with the integration over  zero modes of the 1-loop determinant:
% Z_{\text{2d},\text{ell}}^{[N,\,l],(\vec{k})} 
%\begin{equation}
%\label{eq:1-loop-2d-glsm}
%\begin{split}
%\mathcal{I}_{\{k_1, k_2,\cdots,k_N\}}^{\{n_1, n_2,\cdots,n_N\}} = &
%\dfrac{1}{k_1!\ldots k_{N}!} \prod_{i=1}^N \prod_{\ell=1}^{n_i} \frac{\theta(m_\ell^{(i)} - (i-1)m)}{\eta} \oint \left[ \prod_{i=1}^{N}\prod_{s=1}^{k_i} \dfrac{d \phi_s^{(i)}}{2\pi i} \right]
%\prod_{i=1}^{N} \left( \dfrac{\eta^3 \theta(2\epsilon_+)}{\theta(\epsilon_1)\theta(\epsilon_2)} \right)^{k_i} \\
%& \prod_{i=1}^{N} \prod_{s \neq t}^{k_i} \dfrac{\theta(\phi_s^{(i)} - \phi_t^{(i)})\theta(\phi_s^{(i)} - \phi_t^{(i)} + 2\epsilon_+)}{\theta(\phi_s^{(i)} - \phi_t^{(i)} + \epsilon_1)\theta(\phi_s^{(i)} - \phi_t^{(i)} + \epsilon_2)} \\ 
%& \prod_{i=1}^{N} \prod_{s=1}^{k_i} \dfrac{\theta( (2-i)m + \phi_s^{(i)})\theta(i m -\phi_s^{(i)})}{\theta(\epsilon_+ + (1-i)m + \phi_s^{(i)})\theta(\epsilon_+ - (1-i)m - \phi_s^{(i)})} \\
%& \prod_{i=1}^{N} \prod_{s=1}^{k_i} \prod_{t=1}^{k_{i+1}}
%\dfrac{\theta(\phi_{t}^{(i+1)} - \phi_{s}^{(i)} + \epsilon_-) \theta(-\phi_{t}^{(i+1)} + \phi_{s}^{(i)} + \epsilon_-)}{\theta(\phi_{t}^{(i+1)} - \phi_{s}^{(i)} - \epsilon_+) \theta(-\phi_{t}^{(i+1)} + \phi_{s}^{(i)} - \epsilon_+)} \\
%& \prod_{i=1}^N \prod_{s=1}^{k_i} \prod_{\ell = 1}^{n_i} \dfrac{\theta(\phi_s^{(i)} - m_{\ell}^{(i)} + \epsilon_-)\theta(- \phi_s^{(i)} + m_{\ell}^{(i)} + \epsilon_-)}{\theta(\phi_s^{(i)} - m_{\ell}^{(i)} -\epsilon_+)\theta(- \phi_s^{(i)} + m_{\ell}^{(i)} - \epsilon_+)}.
%\end{split}
%\end{equation}
\begin{align} \label{eq:1-loop-2d-glsm}
& \mathcal{I}_{\{k_1,\cdots,k_{N-1}\}}^{\{n_1,\cdots,n_{N-1}\}}
= \dfrac{1}{\prod_{i=1}^{N-1}k_i!} \oint \left[ \prod_{i=1}^{N-1}\prod_{s=1}^{k_i} \dfrac{d \phi_s^{(i)}}{2\pi i} \right]
Z_{k_1 \ldots k_{N-1}}^{\text{NS5}}(\phi, \epsilon_{1,2})
Z_{k_1 \ldots k_{N-1}}^{\text{D6}}(\phi, m, \epsilon_{1,2})
\prod_{i=1}^{N-1} \prod_{\ell = 1}^{n_i} 
Z_{k_i}^{\text{D4'}}(\phi^{(i)}, m^{(i)}_{\ell}, \epsilon_{1,2}), 
\end{align}
where (for $\theta(z) = \theta_1(\tau, \frac{z}{2\pi i})$)
\begin{equation}
\begin{split} \label{eq:lll}
Z_{k_1 \ldots k_{N-1}}^{\text{NS5}}(\phi, \epsilon_{1,2}) = & 
\prod_{i=1}^{N-1} \left( \dfrac{\eta^3 \theta(2\epsilon_+)}{\theta(\epsilon_1)\theta(\epsilon_2)} \right)^{k_i}
\prod_{i=1}^{N-1} \prod_{s \neq t}^{k_i} \dfrac{\theta(\phi_s^{(i)} - \phi_t^{(i)})\theta(\phi_s^{(i)} - \phi_t^{(i)} + 2\epsilon_+)}{\theta(\phi_s^{(i)} - \phi_t^{(i)} + \epsilon_1)\theta(\phi_s^{(i)} - \phi_t^{(i)} + \epsilon_2)} 
\\
& \prod_{i=1}^{N-1} \prod_{s=1}^{k_i} \prod_{t=1}^{k_{i+1}}
\dfrac{\theta(\phi_{t}^{(i+1)} - \phi_{s}^{(i)} + \epsilon_-) \theta(-\phi_{t}^{(i+1)} + \phi_{s}^{(i)} + \epsilon_-)}{\theta(\phi_{t}^{(i+1)} - \phi_{s}^{(i)} - \epsilon_+) \theta(-\phi_{t}^{(i+1)} + \phi_{s}^{(i)} - \epsilon_+)}, \\
% Z_{k_1 \ldots k_{N-1}}^{\text{D6}}(\phi, m, \epsilon_{1,2}) = & \prod_{i=1}^{N-1} \prod_{s=1}^{k_i} \dfrac{\theta( (1-i)m + \phi_s^{(i)})\theta((1+i) m -\phi_s^{(i)})}{\theta(\epsilon_+ + i m + \phi_s^{(i)})\theta(\epsilon_+ - i m - \phi_s^{(i)})}, \\
Z_{k_1 \ldots k_{N-1}}^{\text{D6}}(\phi, m, \epsilon_{1,2}) = & \prod_{i=1}^{N-1} \prod_{s=1}^{k_i} \dfrac{\theta( m + (1-i)m + \phi_s^{(i)})\theta( m - (1- i) m -\phi_s^{(i)})}{\theta(\epsilon_+ + (1-i) m + \phi_s^{(i)})\theta(\epsilon_+ - (1-i) m - \phi_s^{(i)})}, \\
Z_{k_i}^{\text{D4'}}(\phi^{(i)}, m^{(i)}_{\ell}, \epsilon_{1,2}) = &
\frac{\theta(m_\ell^{(i)} - i m)}{\eta}
\prod_{s=1}^{k_i} \dfrac{\theta(\phi_s^{(i)} - m_{\ell}^{(i)} + \epsilon_-)\theta(- \phi_s^{(i)} + m_{\ell}^{(i)} + \epsilon_-)}{\theta(\phi_s^{(i)} - m_{\ell}^{(i)} -\epsilon_+)\theta(- \phi_s^{(i)} + m_{\ell}^{(i)} - \epsilon_+)}.
\end{split}
\end{equation}
The next step is to integrate over all saddle points, parametrized by eigenvalues $\phi$ of the $\prod_{i=1}^{N-1} U(k_i)$ gauge  connection. This zero-mode integral becomes a contour integral over the space of eigenvalues, whose integral contour is specified by the Jeffrey-Kirwan residue operation \cite{Benini:2013nda,Benini:2013xpa}. The selection of individual residue depends on an auxiliary vector $\eta$, while the final residue sum is independent of it.

It is known that in the special case of $n_1 = \cdots = n_{N-1} = 0$, the Jeffrey-Kirwan residues are classified by $(N-1)$-tuples of Young tableaux \cite{Nekrasov:2012xe,Gadde:2015tra} with the choice of $\eta = (1,1,\cdots,1)$. Each residue corresponds to a possible configuration of the Young tableaux $(Y_1,\cdots,Y_{N-1})$ with $|Y_i| = k_i$. A box $s =(m,n)$ in the $i$'th Young tableau $Y_i$ represents the following pole location for an integral variable $\phi^{(i)}$:
\begin{equation}
\label{eq:pole-2d-YD}
\phi^{(i)} = (i-1)m -\epsilon_+ - (n-1)\epsilon_1 - (m-1)\epsilon_2.
\end{equation}
The classification of JK residues becomes more complicated after introducing the $\prod_{i=1}^{N-1} U(n_i)$ flavor symmetry of D4$'$-branes. We must include new types of residues, being associated to the poles at
\begin{align}
	\label{eq:pole-2d-comp}
	\phi^{(i)} = \epsilon_+ + m_{\ell = 1, \cdots n_i}^{(i)} \qquad\text{ and }\qquad \phi^{(i \pm 1)} - \phi^{(i)} = \epsilon_+ 
\end{align}
in addition to the original ones \eqref{eq:pole-2d-YD}. For example, with $N=2$, all residues are labeled by the Young tableau $Y_1$ and the $n_1$-dimensional vector $\mathbf{u}$ such that $|Y_1|  = k_1 - |\mathbf{u}|$, where each component of the vector $\mathbf{u}$ must be either $0$ or $1$. $|\mathbf{u}|$ denotes the sum of all components of the vector $\mathbf{u}$. A non-zero, $\nu$'th component of $\mathbf{u}$ encodes an integral variable $\phi^{(1)}$ to take the following value:
\begin{align}
	\phi^{(1)} = \epsilon_+ + m_\nu^{(1)}.
\end{align}
The associated residues are obtained by substituting these values to the integral measure \eqref{eq:1-loop-2d-glsm}. Similarly for general $N$, one obtains the elliptic genus $\mathcal{I}_{\{k_1,\cdots,k_{N-1}\}}^{\{n_1,\cdots,n_{N-1}\}}$ of self-dual strings  by summing over all possible JK residues associated to \eqref{eq:pole-2d-YD} and \eqref{eq:pole-2d-comp}.

\subsection{Relation to 5d Wilson loops}
\label{subsec:5d6dcompare}

As the individual elliptic genus represents the BPS states for a given configuration of self-dual strings, the full 6d partition function must be written as their sum.
We note that for a given configuration $(n_1,\cdots,n_{N-1})$ of D4$'$-branes, all the elliptic genera with different winding numbers share a common factor of $Z_{\text{D6-D4}'}^{\{n_1, \cdots, n_{N-1}\}} = \prod_{i=1}^{N-1} \prod_{\ell=1}^{n_i}{\theta(m^{(i)}_\ell - i m)}/{\eta}$ which is the 1-loop determinant contribution of D6-D4$'$ bifundamental Fermi multiplets. As explained at the end of Section~\ref{subsec:2dglsm}, the D6-D4$'$ fermions are decoupled from the interacting IR dynamics. Therefore, we divide the elliptic genera $\mathcal{I}_{\{k_1,\cdots,k_{N-1}\}}^{\{n_1,\cdots,n_{N-1}\}} $ by $Z_{\text{D6-D4}'}^{\{n_1, \cdots, n_{N-1}\}}$. The grand canonical partition is then given by  
%Therefore, we define the following 6d grand canonical partition function, after dividing individual elliptic genera by the common factor $Z_{\text{D6-D4}'}^{\{n_1, \cdots, n_{N-1}\}}$:
\begin{align}
\label{eq:Z6d2d}
	Z_{6d/2d}^{\{n_1, \cdots,n_{N-1}\}} =  \mathcal{I}_0^{N} \cdot \left(\sum_{k_1, \ldots, k_{N-1} \geqslant 0}  \mathfrak{n}_1^{k_1}\mathfrak{n}_2^{k_2}  \cdots \mathfrak{n}_{N-1}^{k_{N-1}} \cdot \mathcal{I}_{\{k_1,\cdots,k_{N-1}\}}^{\{n_1,\cdots,n_{N-1}\}} / Z_{\text{D6-D4}'}^{\{n_1, \cdots, n_{N-1}\}}\right)
\end{align}
 $\mathfrak{n}_{1,\ldots,N-1}$ denote the winding number fugacities for $(N-1)$ different types of self-dual strings. 
The overall multiplicative factor $\mathcal{I}_0^{N}$ captures pure momentum states that are decoupled from stringy modes at low energy. It takes the following form:
\begin{align}
	\mathcal{I}_0^{N} = \text{PE}\left[N\frac{\text{sh}(m \pm \epsilon_- )}{\text{sh}(\epsilon_+ \pm \epsilon_-)}\frac{q}{1-q}\right] = \text{PE}\left[N\frac{t(\mu + \mu^{-1}- u-u^{-1})}{(1-tu)(1-tu^{-1})}\frac{q}{1-q}\right].
\end{align}
$Z_{6d/2d}^{\{n_1, \cdots,n_{N-1}\}}$ is expected to be the Wilson surface operator in M5-branes, whose representation is determined by the arrangement of D4$'$-branes, i.e., $(n_1, n_2,\cdots,n_{N-1})$, as explained in Section~\ref{subsec:2dglsm}. We will see that this expectation becomes true for minuscule representations, while there will be extra subtleties for non-minuscule representations.

We have two different methods of computing the Wilson surface operator in M5-branes. Recall that 5d $\mathcal{N}=1^*$ $U(N)$ gauge theory has the UV completion with non-perturbative corrections, as 6d $(2,0)$ SCFT on $S^1$ describing  a stack of $N$ parallel M5-branes \cite{Douglas:2010iu,Lambert:2010iw,Cordova:2015vwa}. The 5d SYM instantons are the Kaluza-Klein momentum modes along the compactified circle. Similarly, the W-bosons in the Coulomb branch of 5d SYM are uplifted to the 6d self-dual strings, whose tension is proportional to the distance between a pair of M5-branes. Once we identify the instanton fugacity and Coulomb branch parameters of 5d SYM with the 6d momentum and winding fugacities along $T^2$,
\begin{align}
	q = e^{2\pi i\tau},\quad \mathfrak{n}_1 = \frac{w_1}{w_2},\quad 
	\mathfrak{n}_2 = \frac{w_2}{w_3}, \quad \cdots, \quad \mathfrak{n}_{N-1} = \frac{w_{N-1}}{w_{N}} \qquad \text{(with $w_1w_2\cdots w_{N}=1$)},
\end{align}
the various Wilson lines in 5d $\mathcal{N}=1^*$ SYM, studied in Section~\ref{subsec:adhm-extra}, are naturally interpreted as the 6d Wilson surface observable in M5-branes. More concretely, we propose the following equalities:
\begin{align}
	\label{eq:Z5d6d}
	Z_{\text{pert}}^{U(N)} \cdot \mathcal{W}_{\mathbf{R}}^{U(N)} = \bigg(\prod_{i=1}^{N-1} \mathfrak{n}_{i}^{\lambda_i}\bigg)  \cdot Z_{6d/2d}^{\{n_1,\cdots,n_{N-1}\}} \quad \text{ with } \quad \mathbf{R} = \otimes_{i=1}^{N-1} \left({\textstyle\bigwedge^{N-i}}\right)^{n_i}
\end{align}
where $(\lambda_1, \cdots, \lambda_{N-1})$ denotes the lowest weight of the representation $\mathbf{R}$ in $\alpha$-basis. 
One necessary condition for \eqref{eq:Z5d6d} to hold is that the flavor chemical potentials $m_{\ell}^{(i)}$ must disappear from the final expression of the 6d string elliptic genus $\mathcal{I}_{\{k_1,\cdots,k_{N-1}\}}^{\{n_1,\cdots,n_{N-1}\}}/ Z_{\text{D6-D4}'}^{\{n_1, \cdots, n_{N-1}\}}$. It turns out to be true for the cases with $N' = \sum_{i=1}^{N-1} n_i = 1$, which corresponds to the Wilson surfaces in minuscule representations. For non-minuscule representations, i.e. $N' = \sum_{i=1}^{N-1} n_i > 1$, we will report the prescription for the flavor fugacities $e^{-m_\ell^{(i)}}$, under which \eqref{eq:Z5d6d} holds true in most of the winding sectors.

Since the 5d perturbative index $Z_{\text{pert}}^{U(N)}$ and the $U(1)$ factors $\mathcal{I}_0^{N}$ are commonly shared among the Wilson surfaces in different representations, we find it convenient to normalize \eqref{eq:Z5d6d} by the partition function $Z_{\text{pert}}^{U(N)} \cdot \mathcal{W}_{\mathbf{1}}^{U(N)} = Z_{6d/2d}^{\{0, \cdots,0\}}$ \cite{Haghighat:2013gba}. We will test if the following is true:
\begin{align}
\label{eq:Z5d6d-normalized}
\bigg(\prod_{i=1}^{N-1} \mathfrak{n}_{i}^{-\lambda_i}\bigg)  \cdot\frac{\mathcal{W}_{\mathbf{R}}^{U(N)}}{\mathcal{W}_{\mathbf{1}}^{U(N)}} =  \frac{Z_{6d/2d}^{\{n_1,\cdots,n_{N-1}\}}}{Z_{6d/2d}^{\{0, \cdots,0\}}}   \quad \text{ with } \quad \mathbf{R} = \otimes_{i=1}^{N-1}  \left({\textstyle\bigwedge^{N-i}}\right)^{n_i}.
\end{align}

\paragraph{Minuscule representations} When $N=2$ and $n_1 = 1$, we found an agreement between the 6d BPS partition function \eqref{eq:Z6d2d} and the $U(2)$ fundamental Wilson loop $\mathcal{W}_{\mathbf{2}}^{U(2)}$, by taking the double series expansion in the momentum (instanton) fugacity $q$ and the winding (W-boson) fugacity $\mathfrak{n}_1 = w_1^2$, i.e.,
\begin{align}
\label{eq:Z5d6d-su2fnd}
\mathfrak{n}_1^{1/2} \cdot {\mathcal{W}_{\mathbf{2}}^{U(2)}}/\,{\mathcal{W}_{\mathbf{1}}^{U(2)}} = {Z_{6d/2d}^{\{1\}}}/ \,{Z_{6d/2d}^{\{0\}}}   
\end{align}
verified up to $q^4$ and $\mathfrak{n}_1^{4}$ order. For $N=3$ case, the 6d observables $Z_{6d/2d}^{\{0,1\}}$ and $Z_{6d/2d}^{\{1,0\}}$ must be compared with the $U(3)$ Wilson loops in the $\bigwedge^1 = \mathbf{3}$ and $\bigwedge^2 = \overline{\mathbf{3}}$ representations, respectively. We found
\begin{align}
&\mathfrak{n}_1^{1/3}\mathfrak{n}_2^{2/3}  \cdot{\mathcal{W}_{\mathbf{3}}^{U(3)}}/\,{\mathcal{W}_{\mathbf{1}}^{U(3)}} =  {Z_{6d/2d}^{\{0,1\}}}/ \,{Z_{6d/2d}^{\{0,0\}}} \ ,\qquad \nonumber \\
&\mathfrak{n}_1^{2/3}\mathfrak{n}_2^{1/3}  \cdot {\mathcal{W}_{\overline{\mathbf{3}}}^{U(3)}}/\,{\mathcal{W}_{\mathbf{1}}^{U(3)}} = {Z_{6d/2d}^{\{1,0\}}}/ \,{Z_{6d/2d}^{\{0,0\}}}
\end{align}
verified up to $q^2$ and $\mathfrak{n}_1^{2}\mathfrak{n}_2^{2}$ orders. Similarly for $N=4$ case, up to $q^2$ and $\mathfrak{n}_1^{2}\mathfrak{n}_2^{2}\mathfrak{n}_3^{2}$ orders, we checked  
\begin{align}
	&\mathfrak{n}_1^{1/4} \mathfrak{n}_2^{1/2} \mathfrak{n}_3^{3/4}  \cdot{\mathcal{W}_{\mathbf{4}}^{U(4)}}/\,{\mathcal{W}_{\mathbf{1}}^{U(4)}} = {Z_{6d/2d}^{\{0,0,1\}}}/ \,{Z_{6d/2d}^{\{0,0,0\}}} ,\quad \nonumber \\
    &\mathfrak{n}_1^{3/4} \mathfrak{n}_2^{1/2} \mathfrak{n}_3^{1/4}  \cdot{\mathcal{W}_{\overline{\mathbf{4}}}^{U(4)}}/\,{\mathcal{W}_{\mathbf{1}}^{U(4)}} =  {Z_{6d/2d}^{\{1,0,0\}}}/ \,{Z_{6d/2d}^{\{0,0,0\}}} , \\
	&\mathfrak{n}_1^{1/2}\mathfrak{n}_2^{1}\mathfrak{n}_3^{1/2}  \cdot{\mathcal{W}_{\mathbf{6}}^{U(4)}}/\,{\mathcal{W}_{\mathbf{1}}^{U(4)}} = {Z_{6d/2d}^{\{0,1,0\}}}/ \,{Z_{6d/2d}^{\{0,0,0\}}}.\nonumber
\end{align}

\paragraph{Non-minuscule representations}
On the contrary, the flavor chemical potentials $m_\ell^{(i)}$ do \emph{not} generally disappear from the elliptic genera $\mathcal{I}_{\{k_1,\cdots,k_{N-1}\}}^{\{n_1,\cdots,n_{N-1}\}}/ Z_{\text{D6-D4}'}^{\{n_1, \cdots, n_{N-1}\}}$ for non-minuscule representations, preventing \eqref{eq:Z5d6d-normalized} from being true. This is because the elliptic genera of D2-branes receive contributions not only from the Higgs branch, in which D2-branes realize 6d self-dual strings, but also from the twisted Higgs branch, where D2-branes become separable on D4$'$-branes and move along the $x^{6}, x^7, x^8$ directions.
	Specifically, as D2-branes are away from NS5-branes, the latter configuration is no longer associated to 6d $(2,0)$ SCFTs which describe the worldvolume physics of NS5-branes. Instead, it corresponds to the monopole (string) bubbling sector for 5d 't Hooft surface defects on D4$'$-branes wrapping $\mathbf{R}^3 \times T^2$.
	For example, once we take the $M \rightarrow \epsilon_+$ limit which Higgses D2-D6 multiplets, or equivalently removes the D6-brane, 
	the elliptic genera $\mathcal{I}_{\{k_1,\cdots,k_{N-1}\}}^{\{n_1,\cdots,n_{N-1}\}}/ Z_{\text{D6-D4}'}^{\{n_1, \cdots, n_{N-1}\}}$  reduce to the elliptic uplift of the monopole bubbling indices in \cite{Ito:2011ea,Mekareeya:2013ija,Brennan:2018yuj} where $\text{sh}(x)$ is replaced with $\text{sh}(x) \rightarrow \theta(x)$. 
	
	Here we do not attempt to distinguish the Higgs branch states from the others as in Section~\ref{subsec:adhm-extra}. We instead report the following observation: 
    For most of the winding sectors $(k_1, \cdots, k_{N-1})$, except those with non-trivial monopole bubbling indices in \cite{Ito:2011ea,Mekareeya:2013ija,Brennan:2018yuj}, 
    the relation \eqref{eq:Z5d6d-normalized} holds true after discarding all the BPS states that carry non-trivial $\prod_{i=1}^{N-1} U(n_i)$ flavor charges. More precisely, we take the series expansion of the right-hand side of \eqref{eq:Z5d6d-normalized} in the flavor fugacities $e^{-m_\ell^{(i)}}$ and compare its zero-th order coefficient with the left-hand side of \eqref{eq:Z5d6d-normalized}. We found that 
\begin{align}
\label{eq:Z5d6d-non-min}
\mathfrak{n}_1^{1}\cdot ({\mathcal{W}_{\mathbf{3}}^{U(2)}}+{\mathcal{W}_{\mathbf{1}}^{U(2)}})/\,{\mathcal{W}_{\mathbf{1}}^{U(2)}} &=   {Z_{6d/2d}^{\{2\}}}/ \,{Z_{6d/2d}^{\{0\}}}\ \Big|_{\text{$0$-th}} && \text{ at } q^{0,1,2} \text{ and } \mathfrak{n}_1^{2,3} \text{ orders}\\
\mathfrak{n}_1^{3/2}\cdot ({\mathcal{W}_{\mathbf{4}}^{U(2)}}+{2\mathcal{W}_{\mathbf{2}}^{U(2)}})/\,{\mathcal{W}_{\mathbf{1}}^{U(2)}} &=   {Z_{6d/2d}^{\{3\}}}/ \,{Z_{6d/2d}^{\{0\}}}\ \Big|_{\text{$0$-th}} && \text{ at } q^{0,1} \text{ and } \mathfrak{n}_1^{3} \text{ orders}\nonumber\\
 \mathfrak{n}_1^{2/3}\mathfrak{n}_2^{4/3}  \cdot ({\mathcal{W}_{\mathbf{6}}^{U(3)}}+{\mathcal{W}_{\overline{\mathbf{3}}}^{U(3)}})/\,{\mathcal{W}_{\mathbf{1}}^{U(3)}} &={Z_{6d/2d}^{\{0,2\}}}/ \,{Z_{6d/2d}^{\{0,0\}}}\ \Big|_{\text{$0$-th}} && \text{ at } q^{0,1} \text{ and } \mathfrak{n}_1^{1,2,3},\mathfrak{n}_2^{2,3},\mathfrak{n}_1^{2}\mathfrak{n}_2^{1},\mathfrak{n}_1^{2}\mathfrak{n}_2^{2} \text{ orders}\nonumber\\
\mathfrak{n}_1^{4/3}\mathfrak{n}_2^{2/3}  \cdot ({\mathcal{W}_{\overline{\mathbf{6}}}^{U(3)}}+{\mathcal{W}_{\mathbf{3}}^{U(3)}})/\,{\mathcal{W}_{\mathbf{1}}^{U(3)}} &= {Z_{6d/2d}^{\{2,0\}}}/ \,{Z_{6d/2d}^{\{0,0\}}} \ \Big|_{\text{$0$-th}}&& \text{ at } q^{0,1} \text{ and } \mathfrak{n}_1^{2,3},\mathfrak{n}_2^{1,2,3},\mathfrak{n}_1^{1}\mathfrak{n}_2^{2},\mathfrak{n}_1^{2}\mathfrak{n}_2^{2} \text{ orders}\nonumber\\
\mathfrak{n}_1^{1}\mathfrak{n}_2^{1}  \cdot ({\mathcal{W}_{\mathbf{8}}^{U(3)}}+{\mathcal{W}_{\mathbf{1}}^{U(3)}})/\,{\mathcal{W}_{\mathbf{1}}^{U(3)}} &= {Z_{6d/2d}^{\{1,1\}}}/ \,{Z_{6d/2d}^{\{0,0\}}} \ \Big|_{\text{$0$-th}}&& \text{ at } q^{0,1} \text{ and } \mathfrak{n}_1^{1,2,3},\mathfrak{n}_2^{1,2,3},\mathfrak{n}_1^{1}\mathfrak{n}_2^{2},\mathfrak{n}_1^{2}\mathfrak{n}_2^{1},\mathfrak{n}_1^{2}\mathfrak{n}_2^{2} \text{ orders}.\nonumber
\end{align}
For those exceptional sectors, the same truncation makes \eqref{eq:Z5d6d-normalized} true only at $q^0$ order, i.e.,
\begin{align}
\label{eq:Z5d6d-non-min2}
\mathfrak{n}_1^{1}\cdot ({\mathcal{W}_{\mathbf{3}}^{U(2)}}+{\mathcal{W}_{\mathbf{1}}^{U(2)}})/\,{\mathcal{W}_{\mathbf{1}}^{U(2)}} &=   {Z_{6d/2d}^{\{2\}}}/ \,{Z_{6d/2d}^{\{0\}}}\ \Big|_{\text{$0$-th}} && \text{ at } q^{0} \text{ and } \mathfrak{n}_1^{1} \text{ order}\\
\mathfrak{n}_1^{3/2}\cdot ({\mathcal{W}_{\mathbf{4}}^{U(2)}}+{2\mathcal{W}_{\mathbf{2}}^{U(2)}})/\,{\mathcal{W}_{\mathbf{1}}^{U(2)}} &=   {Z_{6d/2d}^{\{3\}}}/ \,{Z_{6d/2d}^{\{0\}}}\ \Big|_{\text{$0$-th}} && \text{ at } q^{0} \text{ and } \mathfrak{n}_1^{1,2} \text{ orders}\nonumber\\
 \mathfrak{n}_1^{2/3}\mathfrak{n}_2^{4/3}  \cdot ({\mathcal{W}_{\mathbf{6}}^{U(3)}}+{\mathcal{W}_{\overline{\mathbf{3}}}^{U(3)}})/\,{\mathcal{W}_{\mathbf{1}}^{U(3)}} &={Z_{6d/2d}^{\{0,2\}}}/ \,{Z_{6d/2d}^{\{0,0\}}}\ \Big|_{\text{$0$-th}} && \text{ at } q^{0} \text{ and } \mathfrak{n}_2^{1},\mathfrak{n}_1^{1}\mathfrak{n}_2^{1},\mathfrak{n}_1^{1}\mathfrak{n}_2^{2} \text{ orders}\nonumber\\
\mathfrak{n}_1^{4/3}\mathfrak{n}_2^{2/3}  \cdot ({\mathcal{W}_{\overline{\mathbf{6}}}^{U(3)}}+{\mathcal{W}_{\mathbf{3}}^{U(3)}})/\,{\mathcal{W}_{\mathbf{1}}^{U(3)}} &= {Z_{6d/2d}^{\{2,0\}}}/ \,{Z_{6d/2d}^{\{0,0\}}} \ \Big|_{\text{$0$-th}}&& \text{ at } q^{0} \text{ and } \mathfrak{n}_1^{1},\mathfrak{n}_1^{1}\mathfrak{n}_2^{1},\mathfrak{n}_1^{2}\mathfrak{n}_2^{1} \text{ orders}\nonumber\\
\mathfrak{n}_1^{1}\mathfrak{n}_2^{1}  \cdot ({\mathcal{W}_{\mathbf{8}}^{U(3)}}+{\mathcal{W}_{\mathbf{1}}^{U(3)}})/\,{\mathcal{W}_{\mathbf{1}}^{U(3)}} &= {Z_{6d/2d}^{\{1,1\}}}/ \,{Z_{6d/2d}^{\{0,0\}}} \ \Big|_{\text{$0$-th}}&& \text{ at } q^{0} \text{ and } \mathfrak{n}_1^{1}\mathfrak{n}_2^{1}  \text{ order}.\nonumber
\end{align}
In spite of the extra D4$'$ states present in $\mathcal{I}_{\{k_1,\cdots,k_{N-1}\}}^{\{n_1, \cdots,n_{N-1}\}}/ Z_{\text{D6-D4}'}^{\{n_1, \cdots, n_{N-1}\}}$, we still regard the above results \eqref{eq:Z5d6d-su2fnd}--\eqref{eq:Z5d6d-non-min2} as a supporting evidence for the validity of the quiver gauge theory introduced in Section~\ref{subsec:2dglsm}. It would be desirable to come up with a way to precisely distinguish the 6d SCFT spectrum from the rest in the elliptic genera.

\subsubsection{qq-characters and $\mathscr{Y}$ observables}

As we did in Section \ref{subsec:adhm-index}, let us conclude our discussion by adding a few remarks about the qq-character interpretation of our elliptic genus \eqref{eq:Z6d2d}. Without D4$'$ branes, the NS5-D6 systems we considered can be thought of as engineering a six-dimensional superconformal $U(1)^{N-1}$ abelian $A_{N-1}$-quiver theory, i.e. a linear quiver theory with $N-1$ $U(1)$ gauge nodes with bifundamental and/or fundamental matter. 
The qq-characters for such a theory, as studied in \cite{Nekrasov:2015wsu,Kimura:2016dys,Kimura:2017auj}, can be classified in terms of tensor product representations $\mathbf{R}$ of the $A_{N-1}$ Lie algebra. In our language they are constructed by adding one or more D4$'$ branes to the NS5-D6 system, where the numbers $(n_1, \ldots, n_{N-1})$ of D4$'$ are related to the tensor product representation $\mathbf{R} = \otimes_{i=1}^{N-1}  \left(\bigwedge^{N-i}\right)^{n_i}$
% \begin{equation}
% \mathbf{R} = (1,0,\ldots,0)^{\otimes n_1} \otimes (0,1,\ldots,0)^{\otimes n_2} \otimes \ldots \otimes (0,0,\ldots,1)^{\otimes n_N} \label{eq:repr}
% \end{equation}
as in \eqref{eq:Z5d6d}.
% , where we are indicating minuscule representations by their Dynkin labels
% \footnote{In the literature, higher qq-characters are often associated to the representation $(n_1,n_2,\ldots, n_N)$ rather than \eqref{eq:repr}; we however think the latter is more natural, both because of their character nature and for their interpretation in terms of Wilson loops/surfaces.} 
More precisely we have
\begin{equation}
\mathcal{I}_0^{-N} Z_{6d/2d}^{\{n_1, \ldots, n_{N-1}\}} Z_{\text{D6-D4}'}^{\{n_1, \cdots, n_{N-1}\}}= \sum_{k_1, \ldots, k_{N-1} \geqslant 0}  \mathfrak{n}_1^{k_1} \ldots \mathfrak{n}_{N-1}^{k_{N-1}}\cdot \mathcal{I}_{\{k_1,\cdots,k_{N-1}\}}^{\{n_1,\cdots,n_{N-1}\}} 
= \mathscr{X}^{(A_{N-1})}_{\mathbf{R}}(m_{\ell}^{(i)}).
\end{equation}
Similarly to what happens in the five-dimensional case, the qq-characters for our six-dimensional quiver theory can be written in terms of correlators of $\mathscr{Y}_i$-operators, although in our setup we have a different $\mathscr{Y}_i$ for each abelian gauge node ($i = 1, \ldots, N-1$). Any correlator involving a number $l_i$ of $\mathscr{Y}_i$-observables (or their inverse) depending on mass parameters $x_j^{(i)}$ for each gauge node can be computed via
\begin{equation}
\prod_{i = 1}^{N-1} \prod_{j = 1}^{l_{i}} \mathscr{Y}_i^{s_j^{(i)}}(x_j^{(i)}) 
= \sum_{k_1, \ldots, k_{N-1} \geqslant 0} \dfrac{\mathfrak{n}_1^{k_1} \ldots \mathfrak{n}_{N-1}^{k_{N-1}}}{k_1!\ldots k_{N-1}!} \oint \left[ \prod_{i=1}^{N-1}\prod_{s=1}^{k_i} \dfrac{d \phi_s^{(i)}}{2\pi i} \right]
Z_{k_1 \ldots k_{N-1}}^{\text{NS5}}%(\phi, \epsilon_{1,2})
Z_{k_1 \ldots k_{N-1}}^{\text{D6}}%(\phi, m, \epsilon_{1,2})
\prod_{i=1}^{N-1} \prod_{j = 1}^{l_i} 
\left[ Z_{k_i}^{\text{D4'}}(x^{(i)}_{j}) \right]^{s_j^{(i)}},
\end{equation}
with $s_j^{(i)} = \pm 1$ and where the functions appearing in the integrand were introduced in \eqref{eq:lll}. For brevity, we suppressed their dependence on $\epsilon_\pm$ and $m$.
The integral is evaluated by summing over the residues at the poles  labeled by $(N-1)$-tuples of the Young tableaux \eqref{eq:pole-2d-YD}; this pole prescription therefore differentiates correlators of $\mathscr{Y}_i$-operators from qq-characters, since the latter receive contributions also from poles of the form \eqref{eq:pole-2d-comp}. 

To illustrate the relation between $\mathscr{X}$ and $\mathscr{Y}_i$'s, let us consider a few examples. When $N=2$ the fundamental (i.e. $n_1 = 1$) qq-character, associated to the fundamental representation $\mathbf{2}$ of $A_1$, can be written as 
\begin{equation}
\mathscr{X}_{\bf 2}^{(A_1)}(m_1^{(1)}) 
= \mathscr{Y}_1(m_1^{(1)}) + \mathfrak{n}_1 \dfrac{P(m_1^{(1)})}{\mathscr{Y}_1(m_1^{(1)} + 2\epsilon_+)},
\end{equation}
where
\begin{equation}
P(\sigma) = \frac{\theta(\sigma + m + \epsilon_+)}{\eta} 
\frac{\theta(\sigma - m + \epsilon_+)}{\eta}.
\end{equation}
The first higher qq-character ($n_1 = 2$), corresponding to the representation $\mathbf{2} \otimes \mathbf{2} = \mathbf{3} \oplus \mathbf{1}$ of $A_1$, reads
\begin{align}
& \mathscr{X}_{\mathbf{3} \oplus \mathbf{1}}^{(A_1)}(m_1^{(1)},m_2^{(1)})
= \mathscr{Y}_1(m_1^{(1)}) \mathscr{Y}_1(m_2^{(1)})  
+ \mathfrak{n}_1 P(m_1^{(1)}) S(m_1^{(1)} - m_2^{(1)}) \dfrac{\mathscr{Y}_1(m_2^{(1)})}{\mathscr{Y}_1(m_1^{(1)} + 2\epsilon_+)} \\
& + \mathfrak{n}_1 P(m_2^{(1)}) S(m_2^{(1)} - m_1^{(1)}) \dfrac{\mathscr{Y}_1(m_1^{(1)})}{\mathscr{Y}_1(m_2^{(1)} + 2\epsilon_+)} 
+ \mathfrak{n}_1^2 P(m_1^{(1)})P(m_2^{(1)}) \dfrac{1}{\mathscr{Y}_1(m_1^{(1)} + 2\epsilon_+)\mathscr{Y}_1(m_2^{(1)} + 2 \epsilon_+)} ,\nonumber
\end{align}
with
\begin{equation}
S(\sigma) = \dfrac{\theta(\sigma + \epsilon_1)\theta(\sigma + \epsilon_2)}{\theta(\sigma)\theta(\sigma + 2\epsilon_+)}.
\end{equation}
When $N = 3$ the fundamental ($n_1 = 0$, $n_2 = 1$) and antisymmetric ($n_1 = 1$, $n_2 = 0$) qq-characters, associated to the $A_2$ representations $\mathbf{3}$ and $\bar{\mathbf{3}}$ respectively, can be decomposed as 
\begin{equation}
\begin{split}
\mathscr{X}_{\mathbf{3}}^{(A_2)}(m_1^{(1)})
& = \mathscr{Y}_1(m_1^{(1)})
+ \mathfrak{n}_1 P_1(m_1^{(1)}) \dfrac{\mathscr{Y}_2(m_1^{(1)} + \epsilon_+)}{\mathscr{Y}_1(m_1^{(1)} + 2\epsilon_+)}
+ \mathfrak{n}_1\mathfrak{n}_2
\dfrac{P_1(m_1^{(1)}) P_2(m_1^{(1)} + \epsilon_+)}{\mathscr{Y}_2(m_1^{(1)} + 3\epsilon_+)}, \\
\mathscr{X}_{\bar{\mathbf{3}}}^{(A_2)}(m_1^{(2)}) 
& = \mathscr{Y}_2(m_1^{(2)})
+ \mathfrak{n}_2 P_2(m_1^{(2)}) \dfrac{\mathscr{Y}_1(m_1^{(2)} + \epsilon_+)}{\mathscr{Y}_2(m_1^{(2)} + 2\epsilon_+)}
+ \mathfrak{n}_1\mathfrak{n}_2
\dfrac{P_2(m_1^{(2)}) P_1(m_1^{(2)} + \epsilon_+)}{\mathscr{Y}_1(m_1^{(2)} + 3\epsilon_+)},
\end{split}
\end{equation}
with
\begin{equation}
%\begin{split}
P_1(\sigma) = \frac{\theta(\sigma + m + \epsilon_+)}{\eta} , 
P_2(\sigma) = \frac{\theta(\sigma - 2m + \epsilon_+)}{\eta}.
%\end{split}
\end{equation}
Similar relations can be found for any $N$ and any representation $\mathbf{R} = \otimes_{i=1}^{N-1}  \left(\bigwedge^{N-i}\right)^{n_i}$ \cite{Kimura:2017auj}.

\section{Concluding remarks}

In this paper we studied the 5d $\mathcal{N} = 1^*$  Wilson loops in a generic $U(N)$ gauge representation, using the ADHM quantum mechanics of D0-D4-D4' branes. The partition function for the grand canonical ensemble of D0-branes is the sum over products between two different Wilson loops, supported on D4 and D4'-branes respectively. We illustrated how to disentangle individual Wilson loops from the full D0-brane partition function in a number of examples. They can be interpreted as the Wilson surface observables in M5-branes, based on the correspondence between 5d maximal $U(N)$ SYM and 6d $(2,0)$ SCFT of $A_{N-1}$--type.  We also studied the Wilson surfaces using the 2d $(0,4)$ quiver gauge theories, which correspond to the UV description of self-dual strings in the presence of a Wilson surface defect. 
Both computations are in agreement for the minuscule Wilson surfaces. On the other hand, the elliptic genera for the non-minuscule representations capture extra BPS states  irrelevant from 6d $(2,0)$ SCFT, displaying only a partial agreement between two independent computations. 

There are a number of open questions that deserve further investigation. It would be important to understand how to distinguish a contribution from extra BPS states, captured in the elliptic genera, supported on D4$'$-branes. It would also be a primary step to study the Wilson surface observables in $(1,0)$ SCFTs, such as E-string theory, in which an emergence of extra states in the elliptic genera is generically observed. Finally, it may be interesting to make a parallel study on the codimension-2 defects in M5-branes, either from 5d SYM perspective or from 6d self-dual strings' perspective.

% In this paper we studied five-dimensional $\mathcal{N} = 1^*$ Wilson loops in generic representation of the $U(N)$ gauge group, as well as their S-dual Wilson surfaces in %the tensor branch of 
% the six-dimensional $\mathcal{N} = (2,0)$ SCFT of $A_{N-1}$ type.
% A UV complete description of these observables can be given in terms of branes intersecting along a codimension four locus; the partition function of the whole brane system contains more information than just the observables we are interested in, but it is still possible to extract from it the desired Wilson loops/surfaces.

% There are a number of open questions that deserve further investigation. First of all, it would be important to understand how to completely remove the contribution of the extra D4$'$ states from Wilson surfaces in non-minuscule representations, in particular from the winding sectors in which ``monopole bubbling'' appears. This would allow us to study Wilson surfaces in more general six-dimensional settings, like E-strings and (??). It may also be interesting to study other kind defects, such as systems of branes intersecting along a codimension two locus.

\vspace{\baselineskip}
\noindent{\bf\large Acknowledgements}

\noindent 
We thank Dongmin Gang, Dongwook Ghim, Chiung Hwang, Sungjay Lee, Piljin Yi, Youngbin Yun, and especially Benjamin Assel, Hee-Cheol Kim, Kimyeong Lee for helpful discussions and comments. 
PA and SK are supported in part by the National Research Foundation of Korea (NRF) Grant 2015R1A2A2A01003124 and 2018R1A2B6004914. The work of PA is also supported in part by Samsung Science and Technology Foundation under Project Number SSTF-BA1402-08 and by the Korea Research Fellowship Program through the National Research Foundation of Korea funded by the Ministry of Science, ICT and Future Planning, grant number 2016H1D3A1938054.

\noindent

% \bibliographystyle{utphys}
% \bibliography{./ref.bib}
\providecommand{\href}[2]{#2}\begingroup\raggedright\endgroup

\end{document}